\documentclass[11pt]{article}
\usepackage{epsfig}
\usepackage{amsfonts}
\usepackage{amsmath}
\usepackage{bbm}
\allowdisplaybreaks[1]

\input pix.sty

\newcommand{\aR}{a^{ }_\rmii{R}}
\newcommand{\aL}{a^{ }_\rmii{L}}

\renewcommand{\eq}{eq.~}
\renewcommand{\eqs}{eqs.~}
\renewcommand{\se}{sec.~}
\renewcommand{\ses}{secs.~}
\renewcommand{\fig}{fig.~}
\renewcommand{\figs}{figs.~}

\newcommand{\Nc}{N_{\rm c}}

\newcommand{\rmO}{{\mathcal{O}}}

\def\lsi{\raise0.3ex\hbox{$<$\kern-0.75em\raise-1.1ex\hbox{$\sim$}}}
\def\gsi{\raise0.3ex\hbox{$>$\kern-0.75em\raise-1.1ex\hbox{$\sim$}}}
\newcommand{\lsim}{\mathop{\lsi}}
\newcommand{\gsim}{\mathop{\gsi}}

\newcommand{\nF}[1]{n_\rmii{F}}
\newcommand{\nB}[1]{n_\rmii{B}}
\newcommand{\rmii}[1]{{\mbox{\tiny\rm{#1}}}}

\newcommand{\im}{\mathop{\mbox{Im}}}

\newcommand{\Tint}[1]{{\hbox{$\sum$}\!\!\!\!\!\!\!\int\,}_{\!\!\!\!\raise-0.9ex\hbox{$\scriptstyle{#1}$}}}
\newcommand{\Tinti}[1]{{{\Sigma}\!\!\!\!\raise0.3ex\hbox{$\int$}_\rmii{${#1}$}}}

\newcommand{\bi}{\begin{itemize}}
\newcommand{\ei}{\end{itemize}}

\newcommand{\hide}[1]{ }
\newcommand{\bsl}[1]{\,\slash\!\!\!\!{#1}\,}
\newcommand{\msl}[1]{\,\slash\!\!\!{#1}\,}

\def\TAsc(#1,#2)(#3,#4,#5)%
{\SetWidth{2.0}\CArc(#1,#2)(#3,#4,#5)\SetWidth{1.0}}
\def\Lwidth{3}

\def\TAgl(#1,#2)(#3,#4,#5){\SetWidth{2.0}\PhotonArc(#1,#2)(#3,#4,#5){\Lwidth}%
{6.283 #3 mul 360 div #4 #5 sub #4 #5 sub mul sqrt mul Tdensity mul}%
\SetWidth{1.0}}
\def\TLgl(#1,#2)(#3,#4){\SetWidth{2.0}\Photon(#1,#2)(#3,#4){\Lwidth}
{#1 #3 sub #1 #3 sub mul #2 #4 sub #2 #4 sub mul add sqrt Tdensity mul}%
\SetWidth{1.0}}
\newcommand{\piC}[1]{\;\parbox[c]{40pt}{\begin{picture}(120,60)(0,-20)
\SetWidth{1.0}\SetScale{0.35} #1 \end{picture}}\;}
\renewcommand{\picc}[1]{\;\parbox[c]{60pt}{\begin{picture}(60,30)(0,0)
\SetWidth{1.0}\SetScale{1.3} #1 \end{picture}}\; }
\def\Lwidth{1.3}

\def\ConnectedA(#1,#2,#3){\piC{#1(60,-15)(75,34,146) #2(60,75)(75,214,326)%
 #3(60,60)(20,190,350)%
 \GBoxc(0,30)(10,10){1} \GBoxc(120,30)(10,10){1}%
  }}
\def\ConnectedB(#1,#2,#3){\piC{#1(60,-15)(75,34,146) #2(60,75)(75,214,326)%
 #3(60,60)(60,0)%
 \GBoxc(0,30)(10,10){1} \GBoxc(120,30)(10,10){1}%
  }}
\def\ConnectedC(#1,#2){\piC{#1(60,-15)(75,34,146) #2(60,75)(75,214,326)%
 \GBoxc(0,30)(10,10){1} \GBoxc(120,30)(10,10){1}%
  }}
\def\ConnectedD(#1,#2){\piC{#1(60,-15)(75,34,146) #2(60,75)(75,214,326)%
 \GBoxc(0,30)(10,10){1} \GBoxc(120,30)(10,10){1}%
 \SetWidth{2.0} 
 \Line(55,55)(65,65)%
 \Line(55,65)(65,55)
  }}

\makeatletter \@addtoreset{equation}{section} \makeatother
\renewcommand{\theequation}{\arabic{section}.\arabic{equation}}
\makeatletter
\renewcommand\section{\@startsection {section}{1}{\z@}%
                                   {-5.5ex \@plus -1ex \@minus -.2ex}
                                   {2.3ex \@plus.2ex}%
                                   {\normalfont\large\bfseries}}
\renewcommand\subsection{\@startsection{subsection}{2}{\z@}%
                                     {-3.25ex\@plus -1ex \@minus -.2ex}%
                                     {1.5ex \@plus .2ex}%
                                     {\normalfont\normalsize\bfseries}}
\renewcommand\thesection {\@arabic\c@section}
\renewcommand\thesubsection   {\thesection.\@arabic\c@subsection}
\renewcommand{\@seccntformat}[1]{%
\csname the#1\endcsname.\hspace{1.0em}}
\makeatother

\begin{document}

\flushbottom

\begin{titlepage}

\begin{flushright}
\vspace*{1cm}
\end{flushright}
\begin{centering}
\vfill

{\Large{\bf
 Improved determination of sterile neutrino
    \\[3mm]
 dark matter spectrum
}} 

\vspace{0.8cm}

J.~Ghiglieri and 
M.~Laine 

\vspace*{0.8cm}

{\em
 Institute for Theoretical Physics, 
 Albert Einstein Center, University of Bern, \\ 
 Sidlerstrasse 5, CH-3012 Bern, Switzerland\\}
 
\vspace*{0.8cm}

\mbox{\bf Abstract}
 
\end{centering}

\vspace*{0.3cm}
 
\noindent
The putative recent indication of an unidentified 3.55 keV X-ray line
in certain astrophysical sources is taken as a motivation for an
improved theoretical computation of the cosmological abundance of 7.1
keV sterile neutrinos. If the line is interpreted as resulting from
the decay of Warm Dark Matter, the mass and mixing angle of the
sterile neutrino are known. Our computation then permits for a
determination of the lepton asymmetry that is needed for producing the
correct abundance via the Shi-Fuller mechanism, as well as for an
estimate of the non-equilibrium spectrum of the sterile neutrinos. The
latter plays a role in structure formation simulations. Results are
presented for different flavour structures of the neutrino Yukawa
couplings and for different types of pre-existing lepton asymmetries,
accounting properly for the charge neutrality of the plasma and
incorporating approximately hadronic contributions.

\vfill

 
\vspace*{1cm}
  
\noindent
August 2015

\vfill

\end{titlepage}

%
\section{Introduction}

Despite considerable phenomenological success, the Standard Model of
particle physics suffers from 
a number of shortcomings: it can explain neither
the presence of neutrino oscillations, 
nor of dark matter, nor of a cosmological baryon asymmetry. 
It is remarkable that all of these problems can in principle
be cured by a simple enlargement
of the theory through three generations of right-handed
neutrinos~\cite{fy}--\cite{numsm1}, 
without changing the underlying theoretical 
principles of gauge invariance and renormalizability. 
Despite its simplicity it remains unclear, of course, 
whether nature makes use of this possibility. 

In the present paper we are
concerned with the dark matter aspect
(for reviews see, e.g.,\ refs.~\cite{numsm2,dm2}). 
For the parameter values that are phenomenologically viable, 
the dark matter sterile neutrinos do not contribute to the 
two observed active neutrino mass differences, so that the
dark matter aspect is partly decoupled from the neutrino
oscillation and baryon asymmetry aspects. The decoupling is
not complete, however: it turns
out that the sterile neutrino dark matter
scenario is tightly constrained~\cite{shifuller}, and only
works if the dynamics responsible for generating a baryon asymmetry
also generates a lepton
asymmetry much larger than the baryon asymmetry~\cite{ms2}, which 
subsequently boosts sterile neutrino dark matter production through an 
efficient resonant mechanism first proposed by 
Shi and Fuller~\cite{sf}. Despite the tight constraints, indications
of a possible observation~\cite{obs1,obs2} demand us
to take this scenario seriously. For our purposes, the complicated dynamics
of ref.~\cite{ms2} simply amounts to the fact that the initial 
state at a temperature of a few GeV possesses certain lepton asymmetries. 

The goal
of the present paper is to refine the analysis of ref.~\cite{shifuller} 
in a number of ways, both theoretically and as far as the numerical
solution of the rate equations is concerned. The resulting spectra
could be used as starting points in structure formation
simulations~(cf.\ e.g.\ refs.~\cite{str0}--\cite{str2} and references therein).

Our plan is the following. 
In \se\ref{se:theory} we review and refine the theoretical 
description of sterile neutrino dark matter production from
a thermalized Standard Model plasma with pre-existing lepton asymmetries. 
The practical implementation of the corresponding 
equations and a strategy for their solution 
are discussed in \se\ref{se:practice}. Numerical solutions
are presented in \se\ref{se:numerics}, and we conclude
with a discussion in \se\ref{se:concl}.

%
\section{General derivation of the rate equations}
\la{se:theory}

Our first goal is to derive a closed set of rate
equations for the sterile neutrino distribution function and for lepton
number densities,\footnote{%
 By ``number densities'' we mean asymmetries, 
 i.e.\ particles minus antiparticles. 
 }
valid both near and far from equilibrium.  
In ref.~\cite{shifuller} such equations were postulated but the 
argument involved phenomenological input, in order to 
account for two types of 
``back reaction''. Our derivation yields 
an outcome that differs slightly from ref.~\cite{shifuller}.   
Moreover, the influence of electric 
charge neutrality of the plasma on the relation 
between lepton number densities and lepton chemical potentials was
not properly accounted for in ref.~\cite{shifuller}. Finally, in order
to obtain a maximal effect, 
the lepton asymmetries of the different flavours were treated as 
equilibrated in ref.~\cite{shifuller}, even though flavour
equilibrium (through active neutrino oscillations) is expected 
to be reached only at temperatures below 10 MeV or so~\cite{eq1,eq2}.
We eliminate all of these simplifications in the present paper. 

We start by defining, in \se\ref{ss:setup}, 
the quantities appearing in the equations. 
In \se\ref{ss:production} an evolution
equation is obtained for the sterile neutrino distribution function. 
In \se\ref{ss:washout} the same is achieved for lepton number
densities, with a right-hand side expressed in terms of lepton chemical 
potentials and sterile neutrino distribution functions. 
The system is completed in \se\ref{ss:relation}, where we relate
lepton chemical potentials 
to lepton densities, taking into account electric charge neutrality
of the Standard Model plasma. 

%
\subsection{Definitions}
\la{ss:setup}

We consider a system consisting of right-handed (sterile) neutrinos and 
Standard Model (SM) particles. The SM particles are assumed to be 
in thermal equilibrium at a temperature $T$. The initial state is 
characterized by non-zero lepton chemical potentials, $\mu_a$, 
associated with different lepton flavours. 
(At low temperatures $T \lsim 10$~MeV the lepton flavours
equilibrate through active neutrino oscillations, so that there is
only a single lepton chemical potential, denoted by $\mu^{ }_L$.) 
The initial state may also
contain an ensemble of sterile neutrinos. 
The two sectors communicate
through Yukawa interactions, parametrized
by neutrino Yukawa couplings. As a result, the distribution function of 
the sterile neutrinos evolves towards its equilibrium form, and
the lepton densities decrease, because lepton number is violated 
in the presence of 
neutrino Yukawa interactions and Majorana masses. 
(However, in practice neither process gets completed within
the lifetime of the Universe.)

Let us denote by $h$ the matrix of neutrino Yukawa couplings. We work out a set
of rate equations to $\rmO(h^2)$. This means that, as soon as we have
factorized a coefficient of $\rmO(h^2)$, the sterile neutrinos can be 
treated as mass eigenstates and free particles, with masses given
by Majorana masses. Sterile neutrinos  
can then be represented by on-shell field operators, 
\ba
 \hat{N}^{ }_{I}(\mathcal{X}) & = & 
 \int \! \frac{{\rm d}^3\vec{k}}{\sqrt{(2\pi)^3 2 E^{ }_{I}}}
 \sum_{\sigma=\pm}
 \biggl[
   \hat{a}^{ }_{I\vec{k}\sigma} u^{ }_{I\vec{k}\sigma}
      e^{-i\mathcal{K}^{ }_I\cdot\mathcal{X}} 
 + 
   \hat{a}^{\dagger}_{I\vec{k}\sigma} v^{ }_{I\vec{k}\sigma}
      e^{i\mathcal{K}^{ }_I \cdot\mathcal{X}} 
 \biggr]
 \;, \la{op_N} \\
 \hat{\bar{N}}^{ }_{I}(\mathcal{X}) & = & 
 \int \! \frac{{\rm d}^3\vec{k}}{\sqrt{(2\pi)^3 2 E^{ }_{I}}}
 \sum_{\sigma=\pm}
 \biggl[
   \hat{a}^{\dagger}_{I\vec{k}\sigma} \bar{u}^{ }_{I\vec{k}\sigma}
      e^{i\mathcal{K}^{ }_I\cdot\mathcal{X}} 
 + 
   \hat{a}^{ }_{I\vec{k}\sigma} \bar{v}^{ }_{I\vec{k}\sigma}
      e^{-i\mathcal{K}^{ }_I\cdot\mathcal{X}} 
 \biggr]
 \;, \la{op_Nbar}
\ea
where
$ 
 E^{ }_{I} \equiv \sqrt{k^2 + M_I^2}
$ 
and 
$
 \mathcal{K}^{ }_I\cdot\mathcal{X} \equiv E^{ }_{I}\, t - \vec{k}\cdot\vec{x}
$.
The index $I$ enumerates the sterile neutrino ``flavours'', and $M_I$
is their mass, which we assume to be real and positive.
The creation and annihilation operators satisfy the 
anticommutation relations 
$
 \{ \hat{a}^{ }_{I\vec{k}\sigma} , \hat{a}^{ \dagger }_{J\vec{p}\tau} \}
 = 
 \delta^{ }_{I\!J} \delta^{ }_{\sigma\tau} \delta^{(3)}(\vec{k-p})
$, 
consistent with 
$
 \{ \hat{N}^{ }_I(t,\vec{x}) , \hat{N}^{\dagger}_J(t,\vec{y})  \} 
 = \delta^{ }_{I\!J} \delta^{(3)}(\vec{x-y})
$.
The on-shell spinors $u,v$ are normalized in a usual way, for instance
${u}^{\dagger }_{I\vec{k}\sigma}  u^{ }_{I\vec{k}\tau} = 
 2 E^{ }_I \delta^{ }_{\tau\sigma}$,
${u}^{\dagger }_{I\vec{k}\sigma}  v^{ }_{I(-\vec{k})\tau}
 = 0 
$, 
$ 
 \sum_\sigma u^{ }_{I\vec{k}\sigma} \bar{u}^{ }_{I\vec{k}\sigma}
 = \msl{\mathcal{K}} + M^{ }_I
$
and
$
 \sum_\sigma v^{ }_{I\vec{k}\sigma} \bar{v}^{ }_{I\vec{k}\sigma}
 = \msl{\mathcal{K}} - M^{ }_I
$. The Majorana nature of the spinors
requires that $u=C\bar v^T$, $v=C\bar u^T$, where 
$C$ is the charge conjugation matrix.

Now, let us define the ensemble occupied by the sterile neutrinos. Their
distribution is described by a density matrix, denoted by $\hat{\rho}^{ }_N$.
We take a rather general ansatz for $\hat{\rho}^{ }_N$, 
assumed however to be diagonal in flavour, momentum, and 
spin indices: 
\be
 \hat{\rho}^{ }_N \equiv Z^{-1}_{N} \exp\Bigl(\int_\vec{k}\sum_{I,\sigma} 
 \mu^{ }_{I\vec{k}} 
 \hat{a}^{\dagger}_{I\vec{k}\sigma} \hat{a}^{ }_{I\vec{k}\sigma} \Bigr)
 \;, \la{rhoN}
\ee
where the function $\mu^{ }_{I\vec{k}}$ is left unspecified except for
being spin-independent, $Z^{ }_{N}$ takes care of overall normalization, and
$
 \int_\vec{k} \equiv \int {\rm d}^3\vec{k} / (2\pi)^3
$.
The density matrix has also been assumed to be $\vec{x}$-independent. 
Note that even though 
the function $\mu^{ }_{I\vec{k}}$ bears some resemblance to 
a chemical potential, it is not identical to one (Majorana
fermions are their own antiparticles and no  
chemical potential can be assigned to them). 

The property originating from $\hat{\rho}^{ }_N$ that we need in 
practice is a phase space distribution function, denoted by
$f^{ }_{I\vec{k}}$, which can be defined as 
\be
 \tr \bigl( \hat{a}^{\dagger}_{I\vec{k}\sigma} \hat{a}^{ }_{J\vec{p}\tau}
 \hat{\rho}^{ }_N \bigr)
 \equiv 
 \delta^{ }_{I\!J} \delta^{ }_{\sigma\tau} \delta^{(3)}(\vec{k-p})
 \, f^{ }_{I\vec{k}}
 \;. \la{def_f}
\ee
Setting the indices equal, this amounts to
\be
 f^{ }_{I\vec{k}} = \frac {(2\pi)^3
 \tr ( \hat{a}^{\dagger}_{I\vec{k}\sigma} \hat{a}^{ }_{I\vec{k}\sigma}
 \hat{\rho}^{ }_N )} {V} 
 \;, \la{operator} 
\ee
where $V$ denotes the volume. We remark that 
the normalization in \eq\nr{operator}
differs from that in ref.~\cite{shifuller} by 
a factor $(2\pi)^3$, and is such that 
the total number density of sterile neutrinos, 
summed over the two spin states, reads
\be
 n^{ }_I = \int \! 
 \frac{{\rm d}^3\vec{k}}{(2\pi)^3}
 \, 2 f^{ }_{I\vec{k}}
 \;.
\ee

\newcommand{\A}{\hat{j}^{ }_a}
\newcommand{\B}{\,\hat{\bar{j}}^{ }_b}
\newcommand{\I}{\int_\mathcal{X} \,e^{i \mathcal{K}\cdot\mathcal{X}}}

Let us then turn to the Standard Model (SM) part. 
For its density matrix we adopt the ansatz
\be
 \hat{\rho}^{ }_\rmii{SM} \equiv
 Z^{-1}_\rmii{SM} \exp
 \Bigl[
 - \frac{1}{T} \Bigl( 
   \hat{H}^{ }_\rmii{SM} - \sum_a \mu^{ }_a \hat{L}^{ }_a
 - \mu^{ }_B \hat{B}  \Bigr) 
 \Bigr]
 \;, \la{rhoSM}
\ee
where $\hat{B}$ is the baryon number operator
(at $T \ll 160$~GeV baryon and lepton numbers are separately
conserved within the SM). 
The lepton number operator associated with flavour $a$ reads
\be
 \hat{L}^{ }_a \equiv
 \int_\vec{x} \Bigl[ 
   \hat{\bar{e}}^{ }_a \gamma^{ }_0 (\aL + \aR) \hat{e}^{ }_a
   + \hat{\bar{\nu}}^{ }_a \gamma^{ }_0 \aL \hat{\nu}^{ }_a
 \Bigr]
 \;. \la{La}
\ee
Here $e_a$ denotes a charged lepton of flavour $a$, 
and $\aL \equiv (1-\gamma_5)/2$, 
$\aR \equiv (1+\gamma_5)/2$  are chiral projectors. 
The left-handed doublet is denoted by 
$\aL \ell_a = \aL (\nu_a, e_a)^T $.

The interaction between SM degrees of freedom and sterile neutrinos
is described by the interaction Hamiltonian
\be
 \hat{H}^{ }_\rmi{int} \equiv
 \sum_{I,a} \int_\vec{x}
   \bigl(
    \hat{\bar{N}}^{ }_I h^{ }_{Ia}\,  \hat{j}^{ }_a 
    +
    \hat{\bar{j}}^{ }_a  h^{*}_{I a}  \hat{N}^{ }_I 
  \bigr)
 \;, \la{Hint}
\ee
where 
$h$ is a $3\times 3$ Yukawa matrix with complex elements $h^{ }_{Ia}$.
The gauge-invariant operators to which 
the sterile neutrinos couple are 
\be
 \A \equiv \hat{\tilde{\phi}}^\dagger \aL \hat{\ell}^{ }_a
 \;, \quad
 \,\hat{\bar{j}}^{ }_a = \hat{\bar{\ell}}^{ }_a \aR \hat{\tilde{\phi}} 
 \;,
\ee
where $\tilde\phi \equiv i \sigma^{ }_2 \phi^*$ is the conjugate 
Higgs doublet. For the moment  
the only property that is needed from SM dynamics is
the spectral function corresponding to these operators, 
\be
 \rho_{ab}(\mathcal{K})  \equiv  
 \I \Bigl\langle \fr12 \bigl\{ \A(\mathcal{X}) , \B(0) \bigr\} \Bigl\rangle
 \;.   \la{spectral}
\ee
In practice we assume this function to be flavour-diagonal, 
i.e.\ $\propto \delta^{ }_{ab}$, however the weight is flavour-dependent 
because charged lepton masses play an important role.\footnote{%
 Note that in the range of temperatures relevant for us, $T \gg 10$~MeV, 
 active neutrino flavour oscillations are not fast enough to play
 a role~\cite{eq1,eq2}, and we can work directly in the interaction basis.
 } 

For describing the dynamics of the coupled system, 
we start from an ``instantaneous'' initial state, 
\be
 \hat{\rho}(0) = \hat{\rho}^{ }_\rmii{SM} \otimes \hat{\rho}^{ }_N
 \;, \la{rho0}
\ee
where the SM part is from \eq\nr{rhoSM} and
the sterile neutrino part is from \eq\nr{rhoN}. In an interaction
picture, the density matrix evolves as
\be
 \hat{\rho}^{ }_\rmii{I}(t)  
 = 
 \hat{\rho}(0)
 - i 
 \int_0^t \! {\rm d}t' \, \bigl[\hat{H}^{ }_\rmii{I}(t'),\hat{\rho}(0)\bigr]
 -
 \int_0^t \! {\rm d}t'
 \int_0^{t'} \!\! {\rm d}t'' \, 
 \bigl[\hat{H}^{ }_\rmii{I}(t'),
 \bigl[\hat{H}^{ }_\rmii{I}(t''),\hat{\rho}(0)\bigr] \bigr]
 + \rmO(h^3)
 \;, \la{rhot}
\ee
where $\hat{H}^{ }_\rmii{I}$ is the Hamiltonian corresponding to \eq\nr{Hint}
in the interaction picture. This evolution causes both the SM and sterile
neutrino density matrices to evolve. For the Standard Model this change
is almost negligible whereas for sterile neutrinos it is of $\rmO(1)$.

In order to obtain differential equations
for the set of observables to be defined, the time $t$ is considered
large compared with time scales of the SM plasma, 
$t \gg 1/(\alpha^2 T)$, where $\alpha$ is a generic 
fine-structure constant. This guarantees that decoherence
takes place in the SM part of the Hilbert space. 
At the same time $t$ should be small enough to avoid 
secular terms. In practice, through an appropriate choice of 
$\hat{\rho}^{ }_\rmii{SM}$ and 
$\hat{\rho}^{ }_N$, we can arrange things in 
a way that the limit $t\to\infty$ can be taken and equilibration is 
correctly accounted for without secular terms (see below). 
An implicit assumption we make is that since the sterile neutrinos
are being produced from a statistical plasma, they are decoherent, 
and their density matrix is assumed to retain the diagonal 
form in \eq\nr{rhoN}; this assumption would surely be 
violated by terms of higher order in $h$. However, given that we
do find a system evolving towards equilibrium, the ansatz 
of \eq\nr{rhoN} can be considered self-consistent. 

%
\subsection{Sterile neutrino production rate}
\la{ss:production}

As a first ingredient, we determine the 
production rate of sterile neutrinos from an
asymmetric plasma. The computation can be carried out
with the formalism of ref.~\cite{hadronic}.\footnote{%
  It could also be carried out
  with the (non-equilibrium) Schwinger-Keldysh formalism. In order to 
  obtain kinetic equations, one may Wigner-transform
  Schwinger-Dyson equations for the
  forward and backward Wightman propagators of the sterile neutrinos 
  (see e.g.~ref.~\cite{sd} for the generic formalism).
 }
From \eq\nr{rhot} 
the density matrix of the system evolves as 
\be
 \dot{\hat{\rho}}^{ }_\rmii{I}(t) = 
 (\mbox{odd powers of $h$})
 - \int_0^t \! {\rm d}t' \, 
 \bigl[ \hat{H}^{ }_\rmii{I}(t), 
 \bigl[ \hat{H}^{ }_\rmii{I}(t'), \hat{\rho}(0) \bigr]\bigr] 
 + \rmO(h^4)
 \;, \la{rho_t}
\ee
where terms containing odd powers of $h$ have been suppressed 
because they are projected out later on. 
The density matrix of the initial state is given by \eq\nr{rho0}.

The observable we are interested in corresponds to the time derivative
of the expectation value of \eq\nr{operator}, normalized like in 
\eq\nr{def_f} (and summed over spin states):
\be
 2\dot{f}^{ }_{I\vec{k}}
 = \frac{(2\pi)^3}{V} \sum_{\sigma} 
 \tr \Bigl( \hat{a}^{\dagger}_{I\vec{k}\sigma} \hat{a}^{ }_{I\vec{k}\sigma}
 \dot{\hat{\rho}}^{ }_\rmii{I} \Bigr) 
 \;.  
\ee
Inserting \eq\nr{rho_t}, we are faced with 4-point functions of 
the creation and annihilation operators, evaluated in an ensemble 
defined by $\hat{\rho}^{ }_N$. Given that the density matrix is 
assumed diagonal in flavour, momentum and spin indices and has an 
effectively Gaussian appearance, the 4-point functions can be 
reduced to 2-point functions:\footnote{%
 Depending on how the computation is organized, some of 
 these relations may not be needed. 
 } 
\ba
 \tr( \hat{a}^\dagger_r \hat{a}^\dagger_k \hat{a}^{ }_p \hat{a}^{ }_q
 \hat{\rho}^{ }_N) & = &
 (\delta^{ }_{rq}\delta^{ }_{kp} - \delta^{ }_{rp}\delta^{ }_{kq} )
 f^{ }_r f^{ }_k
 \;, \\
 \tr( \hat{a}^\dagger_r \hat{a}^{ }_p \hat{a}^\dagger_k \hat{a}^{ }_q
 \hat{\rho}^{ }_N) & = &
 \delta^{ }_{rp}\delta^{ }_{kq} f^{ }_r f^{ }_k
 +  \delta^{ }_{rq}\delta^{ }_{kp}  f^{ }_r (1 - f^{ }_k )
 \;, \\
 \tr( \hat{a}^\dagger_r \hat{a}^{ }_p \hat{a}^{ }_q \hat{a}^\dagger_k
 \hat{\rho}^{ }_N) & = &
 ( \delta^{ }_{rp}\delta^{ }_{kq}
  - \delta^{ }_{rq}\delta^{ }_{kp} )
 f^{ }_r (1- f^{ }_k)
 \;, \\
 \tr( \hat{a}^{ }_p \hat{a}^\dagger_r  \hat{a}^{ }_q  \hat{a}^\dagger_k
 \hat{\rho}^{ }_N) & = &
 \delta^{ }_{rp}\delta^{ }_{kq} 
 ( 1 - f^{ }_r - f^{ }_k + f^{ }_r f^{ }_k ) 
 + 
 \delta^{ }_{rq}\delta^{ }_{kp}  f^{ }_r (1-f^{ }_k)
 \;, \\
 \tr( \hat{a}^{ }_p \hat{a}^\dagger_r \hat{a}^\dagger_k \hat{a}^{ }_q
 \hat{\rho}^{ }_N) & = &
 \delta^{ }_{rp}\delta^{ }_{kq} 
 (1- f^{ }_r) f^{ }_k
 -
 \delta^{ }_{rq}\delta^{ }_{kp}
 f^{ }_r (1- f^{ }_k)
 \;, 
\ea
where the indices incorporate all dependences. 
Subsequently a somewhat tedious analysis, tracing the steps outlined
in ref.~\cite{hadronic}, shows that all terms quadratic in the 
distribution functions drop out. The final result reads 
\ba
 2 \dot{f}^{ }_{I\vec{k}}(t) & = &  
 \sum_a \frac{|h^{ }_{Ia}|^2}{E^{ }_I}
 \Bigl\{
   \bigl[\nF{}(E^{ }_I - \mu^{ }_a) - f^{ }_{I\vec{k}}(t) \bigr] 
   \tr \bigl[ \bsl{\mathcal{K}} \rho^{ }_{aa}(\mathcal{K})\, \aR \bigr]
 \nn 
 && \hspace*{1.6cm} + \,   
   \bigl[\nF{}(E^{ }_I + \mu^{ }_a) - f^{ }_{I\vec{k}}(t) \bigr] 
   \tr \bigl[ \bsl{\mathcal{K}} \rho^{ }_{aa}(- \mathcal{K})\, \aR \bigr]
 \Bigr\}
 \;, \la{production}
\ea
where $\nF{}$ is the Fermi distribution.
On the right-hand side we have replaced $f^{}_{I\vec{k}}(0)$ with
$f^{}_{I\vec{k}}(t)$, i.e. $\hat\rho^{ }_\rmii{I}(0)$ with 
$\hat\rho^{ }_\rmii{I}(t)$ in
eq.~\eqref{rho_t}. This is consistent with our goal of an 
$\mathcal{O}(h^2)$ determination of $f^{}_{I\vec{k}}$, since
the difference between   $\hat\rho^{ }_\rmii{I}(0)$ and 
$\hat\rho^{ }_\rmii{I}(t)$
is of higher order in $h$. As was anticipated at the end of the 
previous section, this prescription removes any secular terms 
from the evolution.
Finally, we remark that, in the limit $f^{ }_{I\vec{k}} \to 0$, 
the result agrees with 
ref.~\cite{shifuller}, whereas in equilibrium, 
corresponding to $\mu^{ }_a =0$, $f^{ }_{I\vec{k}} = \nF{}(E_I)$, 
the production rate vanishes as must be the case. 

%
\subsection{Lepton number washout rate}
\la{ss:washout}

The lepton number defined by \eq\nr{La} is not conserved 
because of the interactions in \eq\nr{Hint}. The  
operator equation of motion reads~\cite{washout}
\be
 \dot{\hat{L}}^{ }_a = 
 \int_\vec{x} i  \Bigl(
  \hat{\bar{N}}^{ } h^{ }_{ }  T^a \hat{j}^{ } - 
  \hat{\bar{j}}^{ } T^a  h^{\dagger}_{ } \hat{N}^{ }
 \Bigr)
 \;, 
\ee
where $(T^a)^{ }_{ij} \equiv \delta^{ }_{ai}\delta^{ }_{aj}$. 
We propose to 
evaluate the expectation value of this operator (taken in the 
interaction picture) in the time-dependent
ensemble described by \eq\nr{rhot}. This time
the leading contribution comes from
the term linear in $\hat{H}^{ }_\rmii{I}$ ($n^{ }_a \equiv L^{ }_a/V$): 
\ba
 \dot{n}^{ }_a \!\! & \equiv & \lim_{t\to \infty} \frac{
 \tr \Bigl[  \dot{\hat{L}}^{ }_a \hat{\rho}^{ }_\rmii{I} (t)\Bigr] 
 }{V}
 \nn 
 & = & 
\lim_{t\to \infty}
 \frac{1}{V} \int_{\vec{x,y}} \int_0^{t} \! {\rm d}t' 
 \Bigl\langle \Bigl[ 
 \bigl(
  \hat{\bar{N}}^{ } h^{ } T^a \hat{j}^{ } - 
  \hat{\bar{j}}^{ } T^a h^{\dagger} \hat{N}^{ }
 \bigr)(\mathcal{X})
   , 
   \bigl(
    \hat{\bar{N}}^{ } h^{ } \hat{j}^{ } 
    +
    \hat{\bar{j}}^{ } h^{\dagger}  \hat{N}^{ } 
  \bigr)(\mathcal{Y})
 \Bigr] \Bigr\rangle + \rmO(h^4) 
 \;, \hspace*{9mm}
\ea
where $\mathcal{X} \equiv (t,\vec{x})$, $\mathcal{Y} \equiv (t',\vec{y})$, 
and $\langle ... \rangle \equiv \tr [(...) \hat{\rho}(0)]$. The SM
part of this expectation value can be expressed in terms of Wightman 
correlators, which in turn can be expressed in terms of the spectral 
function in \eq\nr{spectral}. For the sterile neutrino part, we can
insert the field operators from \eqs\nr{op_N}, \nr{op_Nbar}, and 
eliminate the annihilation and creation operators through 
\eq\nr{def_f}. A tedious but straightforward analysis yields
\ba
 \dot{n}^{ }_a & = &  
 \sum_I \int_\vec{k} \frac{|h^{ }_{Ia}|^2}{E^{ }_I}
 \Bigl\{
   \bigl[f^{ }_{I\vec{k}}(t) - \nF{}(E^{ }_I - \mu^{ }_a)  \bigr] 
   \tr \bigl[ \bsl{\mathcal{K}} \rho^{ }_{aa}(\mathcal{K})\, \aR \bigr]
 \nn 
 && \hspace*{1.6cm} + \,   
   \bigl[\nF{}(E^{ }_I + \mu^{ }_a) - f^{ }_{I\vec{k}}(t) \bigr] 
   \tr \bigl[ \bsl{\mathcal{K}} \rho^{ }_{aa}(- \mathcal{K})\, \aR \bigr]
 \Bigr\} + \rmO(h^4)
 \;, \la{washout}
\ea
where we have employed the prescription described below \eq\nr{production}.
In the limit $f^{ }_{I\vec{k}} \to 0$, the result corresponds to
\eq(2.32) of ref.~\cite{shifuller}.  
On the other hand in equilibrium, 
i.e.\ $\mu^{ }_a =0$, $f^{ }_{I\vec{k}} = \nF{}(E_I)$, 
the washout rate vanishes as must be the case. 

An interesting crosscheck of \eq\nr{washout} 
can be obtained by putting sterile neutrinos in equilibrium
($f^{ }_{I\vec{k}} \to \nF{}(E^{ }_I)$) and by expanding 
lepton number densities to first
order around equilibrium. 
In the limit of small $\mu^{ }_a$, we can re-express
the $\mu^{ }_a$ through the lepton densities
through a susceptibility matrix 
$\Xi^{ }_{ab} = \partial n^{ }_a / \partial {\mu^{ }_b} |_{\mu^{ }_b = 0}$. 
Then we get 
\be
 \dot{n}^{ }_a = - \gamma^{ }_{ab} n^{ }_b + \rmO(h^4, n^2)
 \;, 
\ee
describing how lepton asymmetries disappear (or are ``washed out'')
near equilibrium. The coefficients read
\be
 \gamma^{ }_{ab} = - \sum_I |h^{ }_{Ia}|^2 \int_{\vec{k}}
 \frac{\nF{}'(E^{ }_I)}{E^{ }_I}
 \tr \Bigl\{ \bsl{\mathcal{K}} \Bigl[ \rho^{ }_{aa}(\mathcal{K})
  + \rho^{ }_{aa}(-\mathcal{K}) \Bigr] \aR \Bigr\}\, \Xi^{-1}_{ab}
 \;,
\ee
which after an appropriate adjustment of conventions agrees with 
\eq(4.7) / (29) of ref.~\cite{washout}. 

%
\subsection{Relation of lepton densities and chemical potentials}
\la{ss:relation}

In order to close the set of equations, we need to determine 
the relations between lepton densities and chemical potentials. 
Even though the densities $n_a$ are 
separately conserved within the Standard Model with $h=0$, 
their fluctuations are correlated. The reason is that, because of charge
neutrality, an excess of electrons over positrons is 
compensated for by an excess of antimuons over muons. This implies
that the relation of $\mu_a$ and $n_a$ is non-diagonal. In the present 
section we work it out to leading order in Standard Model couplings, 
at $T \lsim 1$~GeV.
(Recently, similar computations have been extended up to higher orders
in Standard Model couplings
at $T \gsim 160$~GeV~\cite{washout,susc}.)

The desired relations can most conveniently be obtained by first
computing the pressure (i.e.\ minus the grand canonical free energy
density, $p = -\Omega/V$) 
as a function of chemical potentials associated with all
conserved charges. Apart from baryon and lepton numbers, chemical
potentials need to be assigned to gauge charges, in our case the 
electromagnetic U(1)$^{ }_\rmi{em}$ charge ($\equiv \mu^{ }_Q$)
and the weak SU(2)$^{ }_\rmii{L}$ charge ($\mu^{ }_Z$)~\cite{khlebnikov}. 
The chemical potentials assigned to gauge charges correspond to
the fact that the zero components of the associated gauge fields can 
develop an expectation value. In our case, the weak gauge bosons can
be omitted, because their effective potential has a large tree-level
term $\sim \mu_Z^2 v^2$, which implies that $\mu^{ }_Z$ is 
suppressed by $\sim T^2/v^2$ with respect to $\mu^{ }_Q$. Therefore, 
only $\mu_a, \mu^{ }_B$, and $\mu^{ }_Q$ need to be included. 
Given that the chemical potentials are very small compared with 
the temperature ($\mu_a \lsim 0.02T$), 
it is sufficient to determine $p$ up to 
$\rmO(\mu^2)$ in the chemical potentials. 

Omitting exponentially suppressed contributions from the $W^{\pm}$ gauge 
bosons and from the top quark, the pressure can be expressed as
\ba
 p(\mu^{ }_a, \mu^{ }_B, \mu^{ }_Q) & = & 
 p(0) +  \Delta p(\mu^{ }_a, \mu^{ }_B, \mu^{ }_Q)  + \rmO(\mu^4) 
 \;, \\ 
 \Delta p(\mu^{ }_a, \mu^{ }_B, \mu^{ }_Q) & = & 
 \Nc \, 
 \biggl[ 
   \biggl( \frac{\mu^{ }_B + 2 \mu^{ }_Q}{3} \biggr)^2
   \sum_{i=u,c} \chi(m^{ }_i) 
  + 
   \biggl( \frac{\mu^{ }_B - \mu^{ }_Q}{3} \biggr)^2
   \sum_{i=d,s,b} \chi(m^{ }_i) 
 \biggr]
 \nn 
 & + &
 \sum_{a=e,\mu,\tau}
 \Bigl[
   (\mu^{ }_a - \mu^{ }_Q)^2 \chi(m^{ }_{e_a}) + \frac{\mu_a^2}{2}\,\chi(0)
 \Bigr]
 + \rmO(\alpha^3)
 \;, \la{Delta_p}
\ea
where $\Nc = 3$ is a book-keeping variable for hadronic effects, and 
$\chi$ is a (diagonal) ``susceptibility''. 
In the free limit the susceptibility reads
\be
 \chi(m) = 
 2 \int_{\vec{p}} \bigl[ - \nF{}'(E) \bigr]
 = 
 \int \frac{{\rm d}^3\vec{p}}{(2\pi)^3}
 \frac{2 \nF{}(E)[1 - \nF{}(E)]}{T}
 \;, \quad
 E \equiv \sqrt{p^2 + m^2}
 \;. \la{chi_def}
\ee
For vanishing mass this evaluates to 
$
 \chi(0) = T^2/6
$,
whereas for a non-vanishing mass it can be expressed as
\be
 \chi(m) = \frac{m^2}{\pi^2}
 \sum_{n=1}^{\infty} (-1)^{n+1} K^{ }_2\Bigl( \frac{n m }{T} \Bigr)
 \;, 
\ee
where $K^{ }_2$ is a modified Bessel function. (Radiative corrections
to diagonal quark susceptibilities have been computed up to a high order
in the massless limit~\cite{av}, and the same quantities can
also be measured on the lattice, cf.\ e.g.\ refs.~\cite{bmw,rbc}
and references therein.)

Given \eq\nr{Delta_p},  $\mu^{ }_B$ 
is eliminated in favour of the baryon density through 
$n^{ }_B = \partial p / \partial \mu^{ }_B$, yielding
\be
 n^{ }_B=\frac{2\Nc}{9}\Bigl[ \chi^{}_{uc}(\mu^{}_B+2\mu^{}_Q)
 +\chi^{}_{dsb}(\mu^{}_B-\mu^{}_Q) \Bigr]
 \;,
 \label{baryon_density}
\ee
where we have denoted
\be
 \chi^{ }_{abc...} \equiv \sum_{i=a,b,c,...} \chi(m_i)
 \;.
\ee
In the following we neglect $n^{ }_B$ in comparison with lepton densities, 
$n^{ }_B \approx 0$, which fixes $\mu^{}_B$ 
(if $n^{ }_B$ were kept non-zero, 
a Legendre transform  should be carried out
to an ensemble with fixed $n^{ }_B$). 
Charge neutrality is imposed by requiring 
$\partial p / \partial \mu^{ }_Q = 0$. From the resulting
expression, we can obtain lepton densities as functions of
the chemical potentials as 
$
 n_a = \partial p / \partial{\mu^{ }_a}
$.

The solution to the charge neutrality
condition $\partial p / \partial \mu^{ }_Q = 0$ reads
\be
\mu^{}_Q=\frac{\chi^{}_{udscb}}{
      \chi^{ }_{e\mu\tau} 
      \chi^{ }_{udscb}
     + \Nc 
    \chi^{ }_{uc}
    \chi^{ }_{dsb}
   }\sum_{i = e,\mu,\tau} \chi(m^{ }_i) \, \mu^{ }_i
  \;.
   \label{mu_Q_noneq}
\ee
The total lepton density of flavour $a$ is
\be
 n^{ }_a = \chi(0) \, \mu^{ }_a  
   + 2 \chi(m^{ }_a)
   \bigl[ \mu^{ }_a -\mu^{}_Q 
 \bigr] 
 \;. \la{mua}
\ee
The first term accounts for the neutrino density 
\be
 n^{ }_{\nu^{ }_a} = \chi(0) \, \mu^{ }_a 
 \;, \la{na_2}
\ee
whereas the latter term represents
the density $n^{ }_{e_a}$ of charged leptons of flavour $a$. 

It is important to note that because the three lepton 
asymmetries are independent of each other, 
charge conservation can be balanced
by the other lepton flavours. 
For instance, even if $n^{ }_{\nu_\tau} = 0$, so that $\mu^{ }_{\tau} = 0$, 
it is still possible to have $n^{ }_{\tau} \neq 0$ because tau-leptons
couple to $\mu^{ }_Q$ and can thereby neutralize some of the charges
carried by electrons and muons. 

If, however, the different lepton
densities are assumed to equilibrate
(which is physically unlikely at $T > 10$~MeV~\cite{eq1,eq2}), 
the situation changes. We consider this case as well, given that 
it can serve as a useful test case leading to the largest possible
resonance effect~\cite{shifuller}. For equilibrated lepton densities 
we set $\mu^{ }_a = \mu^{ }_L$, $a\in\{1,2,3\}$, so that 
eq.~\eqref{mu_Q_noneq} simplifies to
\be
\mu^{}_Q=\frac{\chi^{}_{e\mu\tau}\chi^{}_{udscb}}{
      \chi^{ }_{e\mu\tau} 
      \chi^{ }_{udscb}
     + \Nc 
    \chi^{ }_{uc}
    \chi^{ }_{dsb}
   }\,\mu^{}_L
  \;, 
   \label{mu_Q_equil}
\ee
and the lepton density is 
$
 n^{ }_L \equiv \sum_a n^{ }_a = 
   3 \chi(0) \mu^{ }_L + 
  2 \chi^{ }_{e\mu\tau}  (\mu^{ }_L - \mu^{ }_Q)
$. 
The neutrino 
and charged lepton asymmetries read 
$
 n^{ }_{\nu^{ }_a} = \chi(0) \, \mu^{ }_L
$  
and
 $n^{ }_{e_a} = 2\chi(m_a) (\mu^{ }_L - \mu^{ }_Q)$, respectively. 
Note that if 
all quark masses are made large compared with temperature, 
or we set $\Nc \to 0$, 
then \eq\nr{mu_Q_equil} implies that $\mu^{ }_Q = \mu^{ }_L$, 
and lepton asymmetry is only carried by neutrinos.  
This is because in the absence of hadronic plasma constituents, 
charge neutrality would be violated if charged leptons
were chemically equilibrated and had non-vanishing
asymmetries. 

%
\section{Practical implementation of the rate equations}
\la{se:practice}

%
\subsection{Summary of the basic setup}

For simplicity, 
we assume that during the period under consideration only one 
sterile neutrino is active, i.e.\  has an interaction rate
comparable with the Hubble rate.\footnote{%
 In the ``$\nu$MSM'', the two heavier sterile
 neutrinos are assumed to have masses 
 in the GeV range~\cite{numsm1,numsm2}, so they are not
 particularly ``heavy''
 at $T\sim 1$~GeV. However their interaction rate peaks
 at higher temperatures, $T \sim 10 - 20$~GeV, 
 cf.\ fig.~3 of ref.~\cite{ms2}, 
 and is very small at $T \lsim 1$~GeV. 
}
Then only
one sterile neutrino
flavour contributes in \eq\nr{washout}. We denote this flavour
by $I=1$. With $\mathcal{K}$ being the associated on-shell four-momentum, 
two rates are defined by
\be
 R^+_{a}(k) \equiv
 \frac{|h^{ }_{1a}|^2 
   \tr[\msl{\mathcal{K}} \rho^{ }_{aa}(\mathcal{K}) \aR]}{E^{ }_1}
 \;, \quad
 R^-_{a}(k) \equiv
 \frac{|h^{ }_{1a}|^2
 \tr[\msl{\mathcal{K}} \rho^{ }_{aa}(-\mathcal{K}) \aR]}{E^{ }_1}
 \;, \la{Rs}
\ee
where $\rho^{ }_{aa}$ is the Standard Model spectral function
from \eq\nr{spectral}. 
Then the basic equations to be solved, \nr{production} and \nr{washout}, 
take the forms ($f^{ }_{k} \equiv f^{ }_{1\vec{k}}$)
\ba
 \dot{f}^{ }_{{k}} & = &
 \fr12 
 \sum_a 
 \Bigl\{ 
   \bigl[\nF{}(E^{ }_1 + \mu^{ }_a) - f^{ }_{{k}} \bigr] R^-_a(k)
  + 
   \bigl[\nF{}(E^{ }_1 - \mu^{ }_a) - f^{ }_{{k}} \bigr] R^+_a(k) 
 \Bigr\} 
 \;, \la{dotfk} \\ 
 \dot{n}^{ }_a & = &  \int_\vec{k}
 \Bigl\{ 
   \bigl[\nF{}(E^{ }_1 + \mu^{ }_a) - f^{ }_{{k}} \bigr] R^-_a(k)
  - 
   \bigl[\nF{}(E^{ }_1 - \mu^{ }_a) - f^{ }_{{k}} \bigr] R^+_a(k)
 \Bigr\} 
 \;, \la{dotna}
\ea
with $\mu^{ }_a$'s and $n^{ }_a$'s related through \eq\nr{mua}. 
The Fermi distributions in \eq\nr{dotfk}, which are always below unity, 
take care of Pauli blocking. 
The two terms on the right-hand sides of these equations 
can be interpreted as originating from reactions involving SM
``particles'' and ``antiparticles''.\footnote{%
 More precisely, the rate $R_a^-$ describes a transition 
 lepton $\leftrightarrow N^{ }_1$ and the rate $R_a^+$ a transition 
 antilepton $\leftrightarrow N^{ }_1$. Time can run in either direction. 
 Eq.~\nr{dotfk} states that sterile neutrinos can be produced from 
 leptons and antileptons. Eq.~\nr{dotna} states that asymmetries between
 lepton and antilepton densities decrease by the difference of the 
 rates felt by leptons and antileptons.  
 } 
In the case of equilibrated active flavours, \eq\nr{dotna} is 
replaced with
\be
 \dot{n}^{ }_{L}  =   \sum_a \int_\vec{k}
 \Bigl\{ 
   \bigl[\nF{}(E^{ }_1 + \mu^{ }_L) - f^{ }_{{k}} \bigr] R^-_a(k)
  - 
   \bigl[\nF{}(E^{ }_1 - \mu^{ }_L) - f^{ }_{{k}} \bigr] R^+_a(k)
 \Bigr\} 
 \;. \la{dotntot}
\ee
Our goal is to solve these equations (generalized to an expanding
background) in the regime
\be
  M_1 = 7.1~\mbox{keV} \;\ll\; 
  E_1 \sim T \sim 200~\mbox{MeV} \;\ll\; 
  v \sim 246~\mbox{GeV}
 \;, \la{inequal}
\ee
where $v$ denotes the Higgs vacuum expectation value.

%
\subsection{Active neutrino properties}

Given that we are deep in the Higgs phase, the Higgs doublet $\tilde{\phi}$
can to a good approximation be replaced by its (gauge-fixed) expectation 
value, cf.\ \eq\nr{inequal}. 
Then the neutrino Yukawa couplings only appear through 
neutrino Dirac masses, defined as
\be
 (M^{ }_\rmii{D})^{ }_{1 a} \equiv \frac{v h^{ }_{1a}}{\sqrt{2}} 
 \;. 
\ee
The spectral function $\rho^{ }_{aa}$ is proportional to the spectral
function of 
an active neutrino of flavour $a$. This assignment refers to 
the weak interaction eigenbasis. 
The (retarded) propagator of an active neutrino
is of the form~\cite{weldon} 
\be
 S^{-1}(-\mathcal{K})
 = \bsl{\mathcal{K}} a + \msl{u} \, 
 \Bigl( b + c + i\, \frac{\Gamma}{2} \Bigr)
 \;, \la{prop}
\ee
and the spectral function is given by 
$
 \rho(\mathcal{K})  = 
 \im S(\mathcal{K} + i u\, 0^+)
$, 
where $u\equiv (1,\vec{0})$ is the four-velocity of the heat bath. 
The function $a$ can to a good approximation be replaced by its
tree-level value $a = 1$~\cite{raffelt}. The functions $b$ and $c$
are often called (thermal and asymmetry induced) matter potentials, 
and $\Gamma$ can be called (up to trivial factors) a thermal width, 
a damping rate, an interaction rate, or an opacity. 
The function $b$ is 
real, even in $\mu^{ }_a$, and odd in $\mathcal{K}$. The function $c$ 
is real, odd in $\mu^{ }_a$, and even in $\mathcal{K}$. The imaginary part
$\Gamma$ can to a good approximation be evaluated
at $\mu^{ }_a = 0$ and is then even in $\mathcal{K}$.
Inserting the form of \eq\nr{prop} into \eq\nr{Rs}, accounting properly for
various sign conventions, and dropping terms which are subleading
in the regime of \eq\nr{inequal}, we obtain
\ba
 R^-_a(k) & \approx &  
 \frac{|M^{ }_\rmii{D}|^2_{1a}\, M_1^2\, \Gamma }
 { [M_1^2 + 2 E^{ }_1(b+c) + (b+c)^2]^2 + E_1^2 \Gamma^2}
 \;, \quad
 R^+_a(k) = 
 \left. R^-_a(k) \right|^{ }_{c\to -c}
 \;.  \la{Rpm}
\ea

Let us now discuss the explicit forms of the functions
appearing in \eq\nr{prop}. In the regime of \eq\nr{inequal} 
the imaginary part $\Gamma$ originates at 2-loop level, 
and has been computed with account of all SM reactions
in ref.~\cite{numsm} (there it was represented 
by the combination $I^{ }_Q  \approx E_1 \Gamma$). It is of 
the form
\be
 \Gamma = G_\rmii{F}^2 T^4 E_1\, \hat{I}^{ }_Q
 \;,  \la{Gamma}
\ee
where $G^{ }_\rmii{F} \equiv g_w^2 / (4\sqrt{2} m_W^2)$ is the Fermi constant, 
and the dimensionless function $\hat{I}^{ }_Q\sim 1$ has been tabulated for 
various momenta, temperatures, and lepton flavours 
on the web page related to ref.~\cite{numsm}. 

The function $b$ originates at 1-loop level and was determined in 
ref.~\cite{raffelt} for $E^{ }_1 \ll m^{ }_W$. It can be expressed as
\ba
 b^{ }_{ } & = & \frac{16 G_\rmii{F}^2 E^{ }_1}{\pi\alpha^{ }_w}
 \Bigl[ \cos^2\!\theta^{ }_w \, \phi(0) + 2 \phi (m^{ }_a) \Bigr]
 \;, \la{baa} \\ 
 \phi(m) & \equiv &  
 \int \frac{{\rm d}^3\vec{p}}{(2\pi)^3}
 \frac{\nF{}(E^{ })}{2 E^{ }}
 \Bigl( \fr43 p^2 + m^2 \Bigr)
 \;, \quad \phi(0) = \frac{7\pi^2 T^4}{360}
 \;, 
\ea
where $\alpha^{ }_w = g_w^2 / (4\pi)$ and $\theta^{ }_w$
is the weak mixing angle. In analogy with \eq\nr{Gamma} we can 
write
\be
 b^{ }_{ } =  G_\rmii{F}^2 T^4 E_1\,  \hat{b}
 \;. 
\ee
Because of the $1/\alpha_w$ factor, 
the dimensionless function $\hat{b}$ is much larger than $\hat{I}^{ }_Q$, 
$\hat{b} \sim 80$. It is plotted in \fig{1}
of ref.~\cite{numsm} for different flavours and temperatures. 

The last ingredient is the function $c$, which incorporates effects
from charge asymmetries. 
This function was also determined in ref.~\cite{raffelt}.
Assuming chiral equilibrium and including all light SM particles,  
we obtain
\ba
\nonumber c &=& \sqrt{2} G^{ }_\rmii{F} 
 \, \Bigl[
  2 n^{ }_{\nu_a} + \sum_{b\neq a} n^{ }_{\nu_b}
 + \Bigl(\fr12 + 2 \sin^2\!\theta^{ }_w\Bigr) n^{ }_{e_a}
 - \Bigl(\fr12 - 2 \sin^2\!\theta^{ }_w\Bigr) \sum_{b\neq a} n^{ }_{e_b}\\
 &&\hspace{1cm} +\Bigl(\frac12-\frac43\sin^2\!\theta^{ }_w\Bigr)
\sum_{i=u,c}n_i
-\Bigl(\frac12-\frac23\sin^2\!\theta^{ }_w\Bigr)
\sum_{i=d,s,b}n_i
\Bigr]
 \;, \la{caa_start}
\ea
where the second line encodes the contribution of quarks,
coming from tadpole diagrams mediated by the $Z$ boson. Because
the latter couples differently  to up- and down-type quarks,
the hadronic contribution contains a part that is not proportional
to $n^{}_B=\frac13\sum_{i=u,d,s,c,b}n_i$ and thus survives even if, 
as we assume, the baryon density vanishes.  The quark densities
read
\be
 n^{ }_u = 
\frac{2 \Nc (\mu^{}_B +2\mu^{}_Q)\chi^{ }_u }{3}
 \;, \quad
 n^{ }_d = 
 \frac{2 \Nc (\mu^{}_B -\mu^{}_Q)\chi^{ }_d}{3}
 \;, \quad
\ee 
and correspondingly for other up- and down-type quarks, respectively. 
Eq.~\eqref{baryon_density}
can be used for expressing $\mu^{}_B$ in terms of
$\mu^{}_Q$, yielding (for $n^{}_B\to0$)
\ba
\nonumber c &=& \sqrt{2} G^{ }_\rmii{F} 
 \, \Bigl[
  2 n^{ }_{\nu_a} + \sum_{b\neq a} n^{ }_{\nu_b}
 + \Bigl(\fr12 + 2 \sin^2\!\theta^{ }_w\Bigr) n^{ }_{e_a}
 - \Bigl(\fr12 - 2 \sin^2\!\theta^{ }_w\Bigr) \sum_{b\neq a} n^{ }_{e_b}\\
 && \hspace{1.2cm}+2(1-2\sin^2\!\theta^{ }_w)
 \frac{\Nc \chi^{}_{uc}\chi^{}_{dsb}}
 {\chi^{}_{udscb}}\,\mu^{}_Q
\Bigr]
 \;. \la{caa}
\ea

We again consider two cases, corresponding to those in \se\ref{ss:relation}.
For {unequilibrated lepton asymmetries}, 
the neutrino asymmetries are different, as given
by \eq\nr{na_2}. 
Then $\mu^{}_Q$ can be read off from eq.~\eqref{mu_Q_noneq}.
The charged lepton densities originate from
the second term in \eq\nr{mua}.
In contrast, for {equilibrated lepton asymmetries},
all neutrino densities are equal, 
$
 n^{ }_{\nu^{ }_a} = \chi(0) \, \mu^{ }_L
$,
whereas $\mu^{}_Q$ can be read off from eq.~\eqref{mu_Q_equil}.

%
\subsection{Expanding background}

In an expanding background, the left-hand sides of \eqs\nr{dotfk}, \nr{dotna}
become
\be
 \dot{f}^{ }_k \to (\partial^{ }_t - H k \partial^{ }_k) f^{ }_k
 \;, \quad
 \dot{n}^{ }_a \to (\partial^{ }_t + 3 H) n^{ }_a
 \;, 
\ee
where $H$ is the Hubble rate, 
$
 H = \sqrt{8\pi e/ (3 m_\rmi{Pl}^2)}
$, 
and $e$ denotes the energy density. 
The inhomogeneous term can be eliminated
from the equation of motion for $f^{ }_k$ by integrating along 
a trajectory of redshifting momentum, 
\be
 k^{ }_T \equiv  k^{ }_* \biggl[\frac{s(T)}{s(T^{ }_*)}\biggr]^{1/3}
 \;, 
\ee
where $s$ is the total entropy density,  
and from that for $n^{ }_a$ by normalizing by $s$, 
\be
 Y_a(T) \equiv \frac{n^{ }_a(T)}{s(T)}
 \;. 
\ee 
It is also 
convenient to integrate in terms of the temperature $T$ rather
than the time $t$. Denoting the final moment of integration 
by $T^{ }_* \equiv 1$~MeV, we get
\ba
 \frac{{\rm d} f^{ }_{k^{ }_T}
 }{{\rm d} \ln(T^{ }_*/T)}
 \!\! & = & \!\! 
 \sum_a \frac{
   \bigl[\nF{}(E^{ }_1 + \mu^{ }_a) - f^{ }_{{k^{ }_T}} \bigr] R^-_a(k^{ }_T)
  + 
   \bigl[\nF{}(E^{ }_1 - \mu^{ }_a) - f^{ }_{{k^{ }_T}} \bigr] R^+_a(k^{ }_T) 
 }{6 H(T) c_s^2(T)}
 \;, \hspace*{6mm} \la{dotfk_2} \\ 
 \frac{{\rm d} Y^{ }_a(T)
 }{{\rm d} \ln(T^{ }_*/T)}
 \!\! & = & \!\!
 \int_{\vec{k}^{ }_T}  
 \frac{
   \bigl[\nF{}(E^{ }_1 + \mu^{ }_a) - f^{ }_{{k^{ }_T}} \bigr] R^-_a(k^{ }_T)
  -
   \bigl[\nF{}(E^{ }_1 - \mu^{ }_a) - f^{ }_{{k^{ }_T}} \bigr] R^+_a(k^{ }_T) 
 }{3 s(T) H(T) c_s^2(T)}
 \;, \la{dotna_2}
\ea 
where $c_s^2$ is the speed of sound squared. Numerical values for 
the thermodynamic functions appearing in these equations 
($e,s,c_s^2$) have been tabulated in ref.~\cite{eos06}.
Note that the right-hand side of \eq\nr{dotfk_2} is even 
in charge conjugation, whereas that of \eq\nr{dotna_2} is odd. 
For the case of equilibrated active flavours, \eq\nr{dotna_2} gets
replaced with ($Y^{ }_L \equiv \sum_a Y^{ }_a$) 
\be
 \frac{{\rm d} Y^{ }_L (T)
 }{{\rm d} \ln(T^{ }_*/T)}
 = 
 \sum_a \int_{\vec{k}^{ }_T}  
 \frac{
   \bigl[\nF{}(E^{ }_1 + \mu^{ }_L) - f^{ }_{{k^{ }_T}} \bigr] R^-_a(k^{ }_T)
  -
   \bigl[\nF{}(E^{ }_1 - \mu^{ }_L) - f^{ }_{{k^{ }_T}} \bigr] R^+_a(k^{ }_T) 
 }{3 s(T) H(T) c_s^2(T)}
 \;. \la{dotn_2}
\ee

%
\subsection{Resonance contributions} 
\la{ss:resonance}

For small $\Gamma$ the rates in \eq\nr{Rpm} resemble Dirac delta-functions, 
if the first factor in the denominator has a zero. For positive $c$, 
this can be the case with the term $R_a^+$. Denoting  
\be
 \mathcal{F}(T) \equiv M_1^2 + 2 E^{ }_1(b-|c|) + (b-|c|)^2
 \;, \la{FT}
\ee
we can then approximate
\be
 R_a^+ \approx
  \frac{|M^{ }_\rmii{D}|^2_{1a}\, M_1^2}{E_1} 
 \frac{I^{ }_Q}
 { \mathcal{F}^2 + I_Q^2}
 \approx 
  \frac{|M^{ }_\rmii{D}|^2_{1a}\, M_1^2\, 
 \pi\, \delta(\mathcal{F}(T))}{E^{ }_1}
 \;, 
 \quad
 I^{ }_Q = E^{ }_1 \Gamma
 \;. \la{Ra_res}
\ee
If $c < 0$, a similar term exists in $R_a^-$; 
it can only exist in one of the terms at a time.  

In general, there are {\em two} resonances to be considered. 
The simplest way to see this is to fix $T$ and consider $\mathcal{F}$
as a function of $E_1$. Indeed the energy dependence
of the variables appearing in \eq\nr{FT} is simple: $b$ is linear
in $E^{ }_1$ whereas $c$ is constant (cf.\ \eqs\nr{baa}, \nr{caa}).
If we write $b = \tilde{b}\, E^{ }_1$, where $\tilde{b} \ll 1$, then 
\be
 \mathcal{F} = \tilde{b}\,(2+\tilde{b}) E_1^2 - 2 E_1 |c| (1+\tilde{b}) + 
 M_1^2 + c^2 
 \;, 
\ee
and zeros exist if $c^2 > \tilde{b}\,(2+\tilde{b}) M_1^2$. 
The zeros are located at 
\be
 E^{ }_{1\pm} \; \equiv \; \frac{|c|(1+\tilde{b}) \pm
 \sqrt{c^2 - \tilde{b}\,(2+\tilde{b}) M_1^2} }{\tilde{b}\, (2+\tilde{b})}
 \;, \la{Epm}
\ee
and the ``Jacobian'' reads $| \partial^{ }_{E_1} \mathcal{F} | = 
{2 \sqrt{c^2 - \tilde{b}\,(2+\tilde{b}) M_1^2}}$
at $E^{ }_1 = E^{ }_{1\pm}$. 

In practice, of course, $I^{ }_Q$ is not infinitesimally small, and
the resonance is not arbitrarily narrow. In this situation resonance
effects interfere with non-resonant contributions. One way to account
for this in a practical numerical solution is sketched in appendix~A.

%
\subsection{Relic density}
\la{ss:relic}

Once the final spectrum $f^{ }_{k_*}$ has been obtained
through the integration of \eqs\nr{dotfk_2}--\nr{dotn_2}, we need to 
relate it to the present-day dark matter energy density. In practice
we choose the lowest temperature of the integration to be 
$T^{ }_* \equiv 1$~MeV, by which time all the source terms have 
switched off, whereas $T^{ }_0$ denotes the present-day temperature
of the cosmic microwave background. Today, the sterile neutrinos 
are non-relativistic, so that the energy density carried by them reads
\be
 \rho^{ }_{1} = M^{ }_1 \int \! \frac{{\rm d}^3\vec{k}_0}{(2\pi)^3}
 \, 2 f^{ }_{k_0}
 \;. \la{rho_1}
\ee
The dark matter energy density can be written as 
($ 
 \Omega^{ }_\rmi{dm} \equiv 
 \rho^{ }_\rmi{dm} / \rho^{ }_\rmi{cr}
$)
\be
 \rho^{ }_\rmi{dm} = \Omega^{ }_\rmi{dm} h^2 \times 
 \frac{\rho^{ }_\rmi{cr}}{ h^2 s(T^{ }_0)} \times s(T^{ }_0)
 \;,  \la{rho_dm}
\ee
where $\rho^{ }_\rmi{cr}$ is the critical energy density and 
$s(T^{ }_0) = 2\,891 / \mbox{cm}^3$ is the current entropy density. 
Making use of the known value of $\rho^{ }_\rmi{cr}$ yields
$
 {\rho^{ }_\rmi{cr}}/ [{ h^2 s(T^{ }_0)}] = 3.65
$~eV. 
Recalling that $\Omega^{ }_\rmi{dm} h^2 = 0.12$ according to Planck 
data~\cite{planck}, and dividing \eq\nr{rho_1} by \eq\nr{rho_dm}, we get
\be
 \frac{\Omega^{ }_1}{\Omega^{ }_\rmi{dm}}
 = \frac{2 M_1}{0.12 \times 3.65\,\mbox{eV}}
 \int \frac{{\rm d}^3\vec{k}_0}{(2\pi)^3}
 \, \frac{f^{ }_{k_0}}{s(T^{ }_0)}
 = 
 6950\times \frac{M^{ }_1}{7.1\,\mbox{keV}} \times
 \int \frac{{\rm d}^3\vec{k}_*}{(2\pi)^3}
 \, \frac{f^{ }_{k_*}}{T^{3}_*}
 \;,  \la{Omega1}
\ee 
where we made use of the facts that 
$
 \int\! {\rm d}^3\vec{k}^{ }_T\, f^{ }_{k_T} / s(T^{ }_{ })
$
is temperature-independent at $T \le T^{ }_*$ and that 
$s(T^{ }_*) \approx 4.67 T_*^3$.
So, given the known $f^{ }_{k_*}$, 
\eq\nr{Omega1} allows us to determine
$\Omega^{ }_1 / \Omega^{ }_\rmi{dm}$.

%
\section{Numerical results}
\la{se:numerics}

\begin{figure}[t]


\centerline{%
 \epsfysize=7.5cm\epsfbox{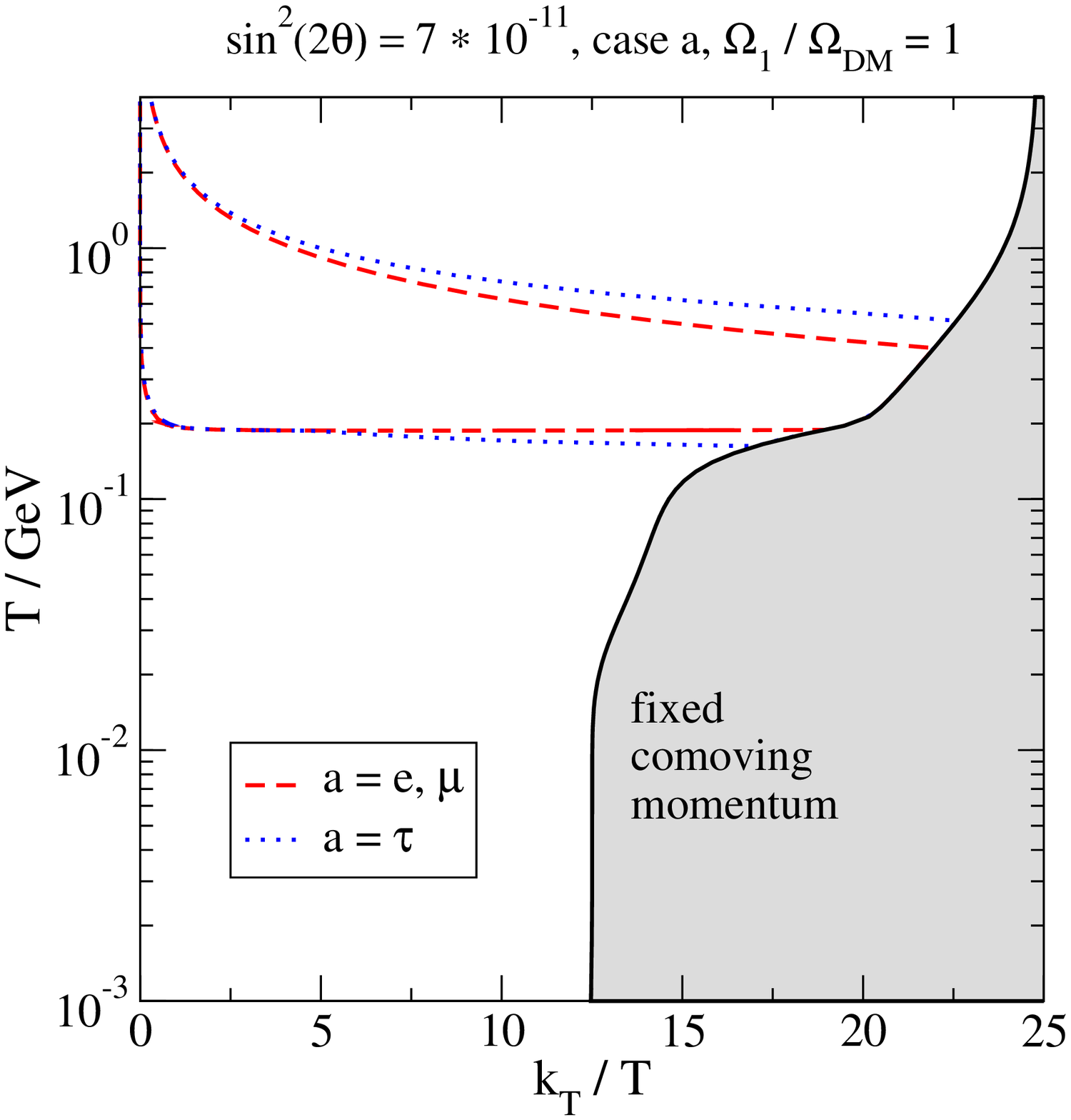}%
~~~\epsfysize=7.5cm\epsfbox{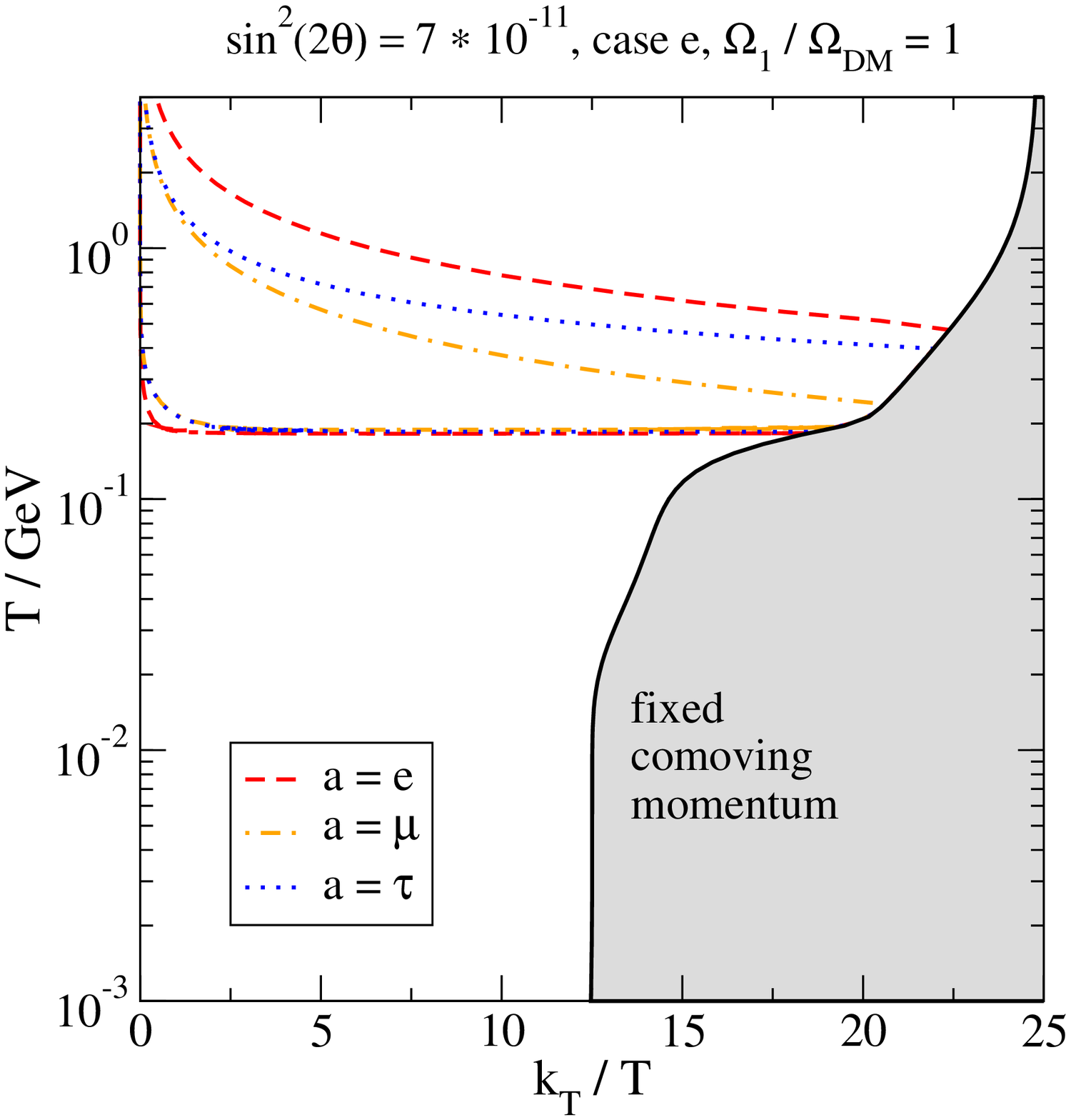}
}

\caption[a]{\small
 Resonance locations in the different flavour channels for 
 case (a) (left) and case (e) (right) 
 [only $a=e$ has a physical effect in these cases 
 because only $h^{ }_{1e}\neq 0$].
 In each case we have set 
 $\sin^2(2\theta) = 7\times 10^{-11}$ and tuned the initial
 asymmetry to the value producing the correct dark matter abundance, 
 cf.\ table~\ref{tab_crit}.  
 We have considered comoving momenta $k^{ }_{T_*} \le 12.5 T^{ }_*$
 at $T^{ }_* = 1$~MeV; the 
 continuous line indicates the upper edge of this range. 
}

\la{fig:resonance}
\end{figure}

\begin{figure}[t]


\centerline{%
 \epsfysize=7.5cm\epsfbox{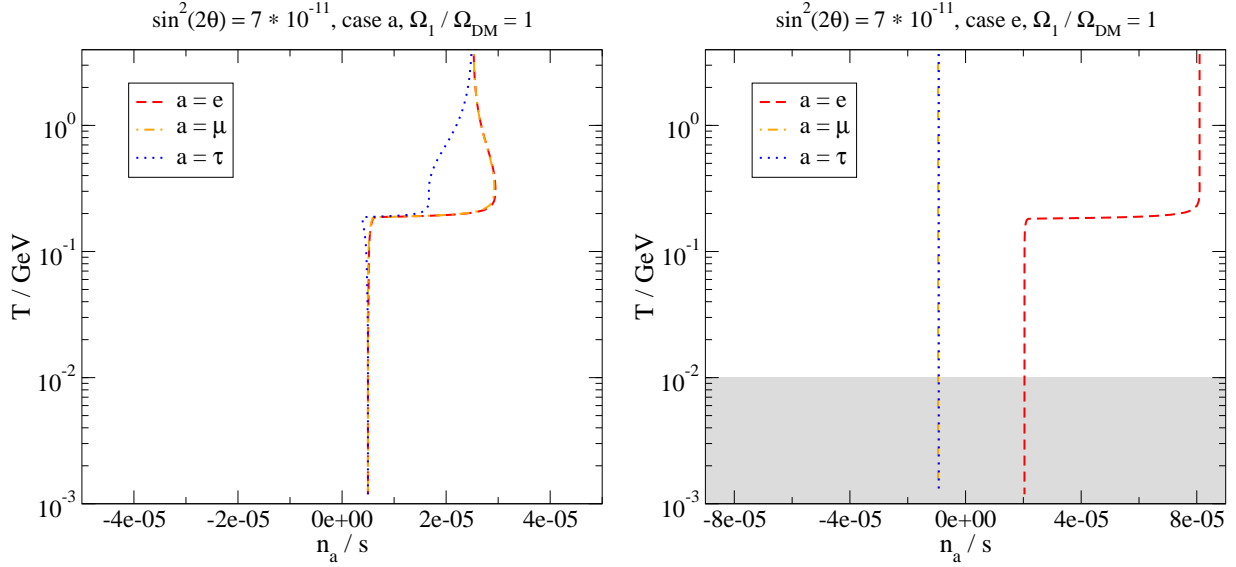}%
~~~\epsfysize=7.5cm\epsfbox{densitye.eps}
}

\caption[a]{\small
 The evolution of the lepton asymmetries $Y^{ }_a$  for 
 case (a) (left) and case (e) (right). The parameters 
 are like in \fig\ref{fig:resonance}. In case (a) 
 $Y^{ }_e, Y^{ }_\mu$ grow initially, even though
 the source terms $R_a^\pm$ are not active yet, because 
 charged $\tau$-leptons cannot carry their share of the 
 asymmetry when $T \ll m^{ }_\tau$ 
 ($Y^{ }_L \equiv \sum_a Y^{ }_a$ is constant). 
 In case (e) such a re-distribution is not possible 
 and $Y^{}_\mu$ and $Y^{}_\tau$ are exactly conserved.
 The values of the initial neutrino asymmetries $n^{ }_{\nu_a}$ are 
 given in table~\ref{tab_crit}; the values of the
 corresponding lepton asymmetries $n^{ }_a = n^{ }_{\nu_a} + n^{ }_{e_a}$
 follow from \eqs\nr{mu_Q_noneq}--\nr{na_2}.
 Lepton asymmetries would be expected to equilibrate
 below $T = 10$~MeV~\cite{eq1,eq2}, in the  
 region shown by a grey band, however the 
 rates $R^\pm_a$ have switched off by then so this has no effect
 on sterile neutrino distributions. 
}

\la{fig:density}
\end{figure}

\begin{figure}[t]
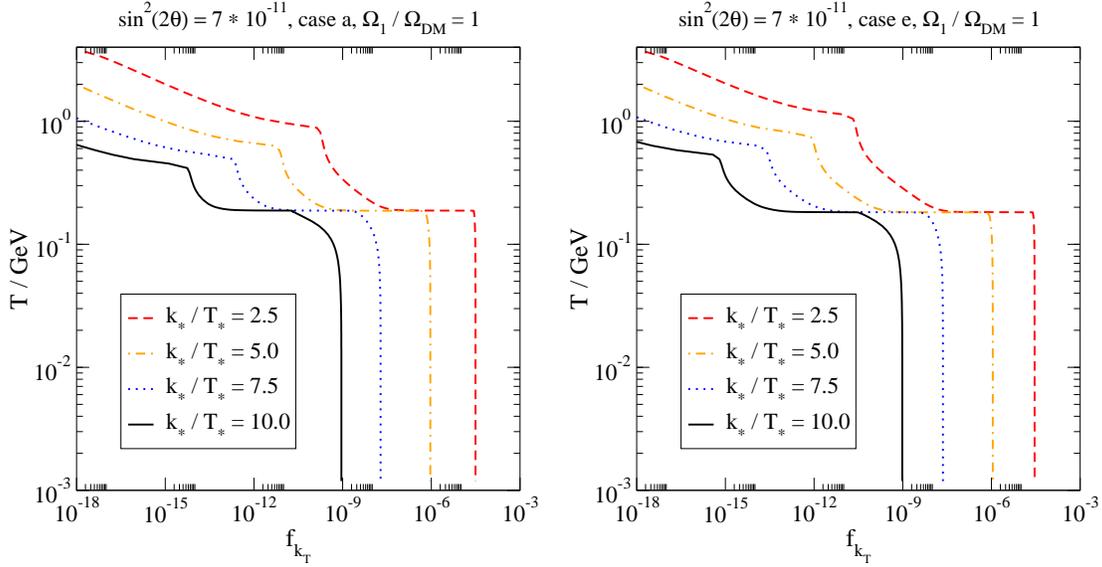



\centerline{%
 \epsfysize=7.5cm\epsfbox{fkTa.eps}%
~~~\epsfysize=7.5cm\epsfbox{fkTe.eps}
}

\caption[a]{\small
 Examples for the 
 evolution of the right-handed neutrino distribution $f^{ }_{k_T}$ for 
 case (a) (left) and case (e) (right), 
 assuming that $f^{ }_{k_T}(T = \mbox{4~GeV}) = 0$.
 The final temperature is $T^{ }_* = 1$~MeV, and 
 $k^{ }_* \equiv k^{ }_{T_*}$ denotes momenta at
 this temperature.  
 The parameters 
 are like in \figs\ref{fig:resonance}, \ref{fig:density}.
 For smallish $k^{ }_*/T^{ }_*$ most of the production
 takes place at the lower resonance temperature 
 (cf.\ \fig\ref{fig:resonance}).
}

\la{fig:fkT}
\end{figure}

\begin{figure}[t]


\centerline{%
 \epsfysize=7.5cm\epsfbox{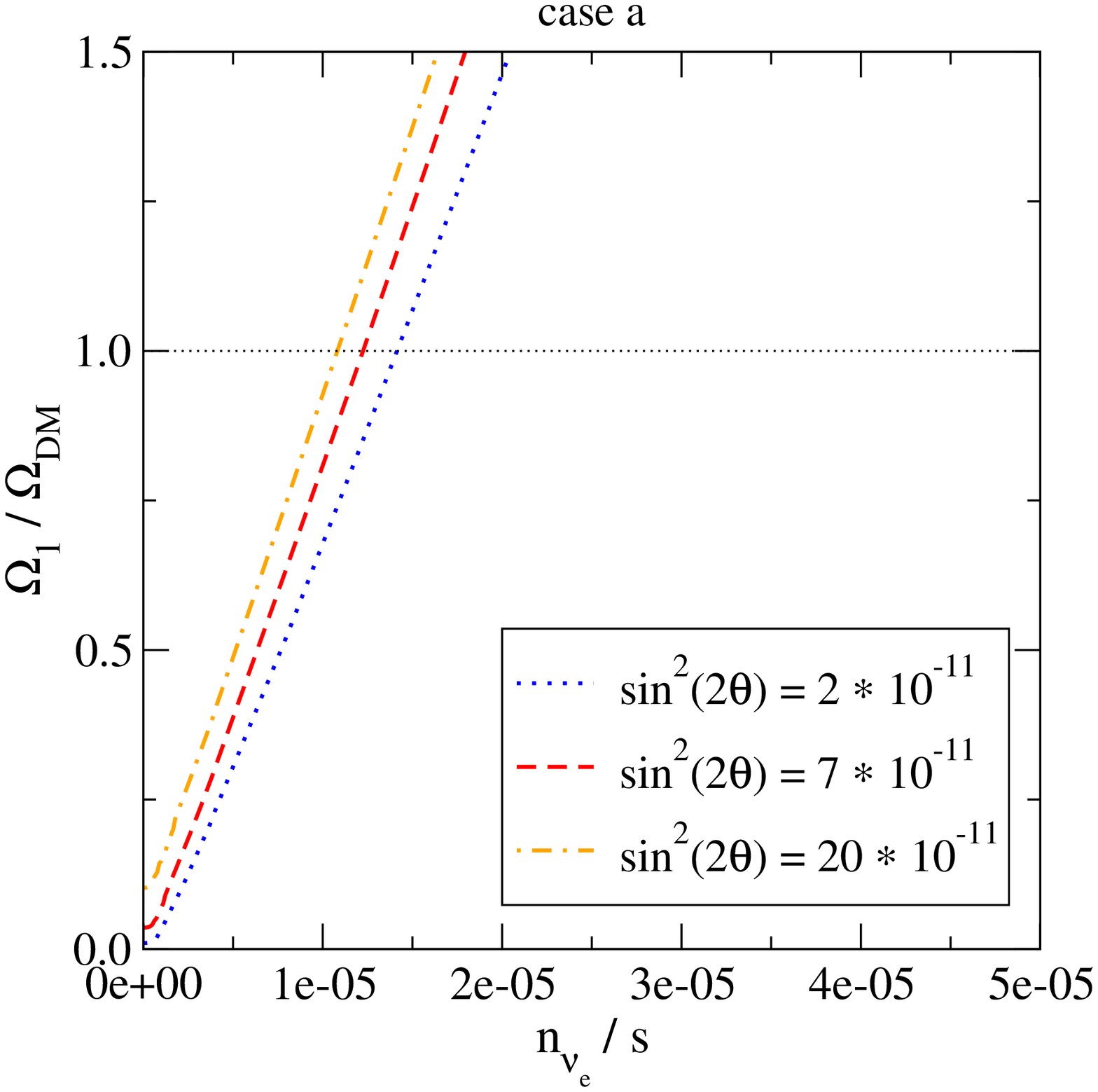}%
~~~\epsfysize=7.5cm\epsfbox{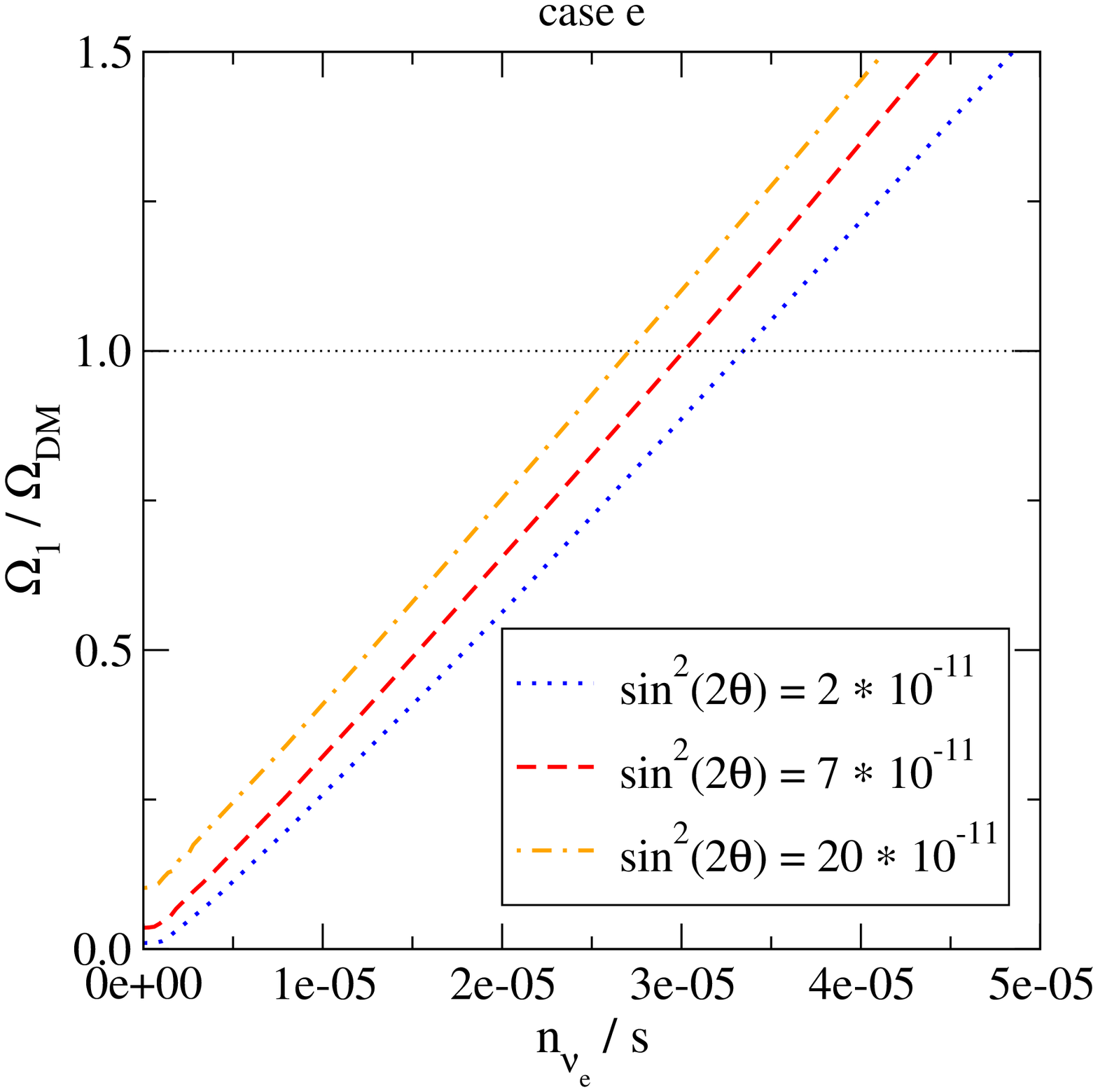}
}

\caption[a]{\small
 The total energy density carried by sterile 
 neutrinos today, normalized to the dark matter density, as a function
 of the initial lepton asymmetry, for 
 case (a) (left) and case (e) (right). The couplings span 
 the range indicated by refs.~\cite{obs1,obs2}.
}

\la{fig:scan}
\end{figure}

\begin{figure}[t]
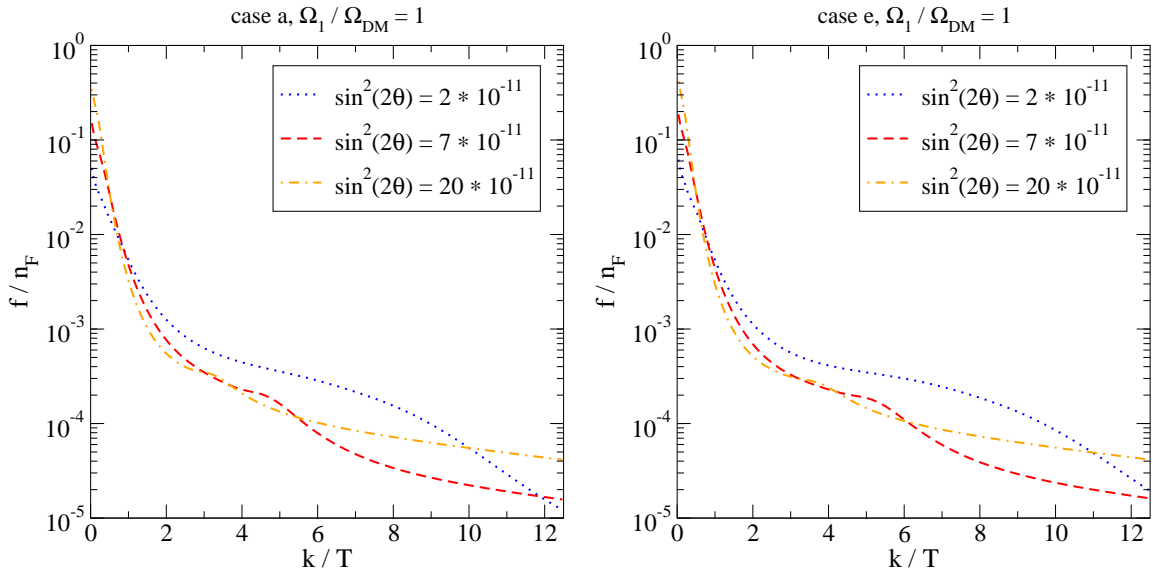



\centerline{%
 \epsfysize=7.5cm\epsfbox{distributiona.eps}%
~~~\epsfysize=7.5cm\epsfbox{distributione.eps}
}

\caption[a]{\small
 The sterile neutrino distribution function
 at $T \le T^{ }_* = 1$~MeV, normalized to the Fermi
 distribution $\nF{}(k^{ }_{ }) = 1/[\exp(k^{ }_{ }/T)+1]$, 
 for the initial lepton asymmetry
 producing precisely the observed dark matter energy density, for 
 case (a) (left) and case (e) (right). Given that $f \ll \nF{}$, 
 sterile neutrinos are far below equilibrium despite their 
 efficient resonant production. 
}

\la{fig:distribution}
\end{figure}

We have integrated \eqs\nr{dotfk_2}--\nr{dotn_2} numerically 
for a number of parameter values, starting at 
$T = T^{ }_\rmi{max}  \equiv 4$~GeV where we assume 
$f^{ }_{k_{T}} = 0$, and stopping at
$T = T^{ }_* \equiv 1$~MeV where all source terms have switched off. 
Subsequently we determine the 
observables defined in \se\ref{ss:relic}
($f^{ }_{k_*}$, $\Omega^{ }_1 / \Omega^{ }_\rmii{DM}$)
for $M^{ }_1 = 7.1$~keV. 
To illustrate
our results, let us focus on the following cases: 
\bi

\item[(a)]
$n^{ }_{\nu_e} = n^{ }_{\nu_\mu} = n^{ }_{\nu_\tau}$ at $T = T^{ }_\rmi{max}$; 
only $h^{ }_{1e}\neq 0$; 
equilibrated active flavours.

\item[(b)]
$n^{ }_{\nu_e} = n^{ }_{\nu_\mu} = n^{ }_{\nu_\tau}$ at $T = T^{ }_\rmi{max}$; 
only $h^{ }_{1e}\neq 0$; 
non-equilibrated active flavours.

\item[(c)]
$n^{ }_{\nu_e} = n^{ }_{\nu_\mu} = n^{ }_{\nu_\tau}$ at $T = T^{ }_\rmi{max}$; 
only $h^{ }_{1\tau}\neq 0$; 
equilibrated active flavours.

\item[(d)]
$n^{ }_{\nu_e} = n^{ }_{\nu_\mu} = n^{ }_{\nu_\tau}$ at $T = T^{ }_\rmi{max}$; 
only $h^{ }_{1\tau}\neq 0$; 
non-equilibrated active flavours.

\item[(e)]
only $n^{ }_{\nu_e} \neq 0 $ at $T = T^{ }_\rmi{max}$; 
only $h^{ }_{1e}\neq 0$; 
non-equilibrated active flavours.

\item[(f)]
only $n^{ }_{\nu_e} \neq 0 $ at $T = T^{ }_\rmi{max}$; 
only $h^{ }_{1\mu}\neq 0$; 
non-equilibrated active flavours.

\item[(g)]
only $n^{ }_{\nu_e} \neq 0 $ at $T = T^{ }_\rmi{max}$; 
only $h^{ }_{1\tau}\neq 0$; 
non-equilibrated active flavours.

\item[(h)]
only $n^{ }_{\nu_\tau} \neq 0 $ at $T = T^{ }_\rmi{max}$; 
only $h^{ }_{1e}\neq 0$; 
non-equilibrated active flavours.

\item[(i)]
only $n^{ }_{\nu_\tau} \neq 0 $ at $T = T^{ }_\rmi{max}$; 
only $h^{ }_{1\mu}\neq 0$; 
non-equilibrated active flavours.

\item[(j)]
only $n^{ }_{\nu_\tau} \neq 0 $ at $T = T^{ }_\rmi{max}$; 
only $h^{ }_{1\tau}\neq 0$; 
non-equilibrated active flavours.

\ei
Let us reiterate that 
in the case of equilibrated active flavours, one would
have to assume active
neutrino oscillations to proceed much faster than the processes considered
in the present paper, which is unlikely to happen 
at $T > 10$~MeV~\cite{eq1,eq2}. Nevertheless we display the results
in order to allow for a comparison with ref.~\cite{shifuller},
to be performed in \se\ref{se:concl}.

The initial state
is parametrized by the neutrino asymmetry 
normalized to the entropy density, $n^{ }_{\nu_a}/s$.
The mixing angles are 
parametrized through
\be
 \sin^2(2\theta) \equiv \sum_{a=e,\mu,\tau} 4 \theta_{1a}^2
 \;, \quad
 \theta_{1a}^2 \equiv \frac{|M^{ }_\rmii{D}|^2_{1a}}{M_1^2}
 \;,
\ee
which is the combination 
that appears in the (inclusive) decay rate of 
sterile neutrinos to an active neutrino and a photon.
We consider the value 
$\sin^2(2\theta) \approx 7\times 10^{-11}$ 
mentioned in ref.~\cite{obs1} and the limits of 
$\sin^2(2\theta) \sim (2 - 20)\times 10^{-11}$
from ref.~\cite{obs2}.  
Confining effects are modelled
through the phenomenological replacement
$\Nc \to N^{ }_\rmi{c,eff}$ as suggested in ref.~\cite{numsm}. 
(In ref.~\cite{numsm} it was checked that this recipe is 
consistent with Chiral Perturbation Theory at low $T$; 
unfortunately Chiral Perturbation Theory is not applicable
at $T \gsim 100$~MeV.)

In \fig\ref{fig:resonance}, the two resonance locations (in each channel)
are shown for the cases (a) and (e). In \fig\ref{fig:density}, the evolution
of the densities $Y^{ }_a$ is shown, and in \fig\ref{fig:fkT} the same
is done for the distribution function $f^{ }_{k_T}$. The ratio
$\Omega^{ }_1 / \Omega^{ }_\rmii{DM}$ from 
\eq\nr{Omega1} is illustrated in \fig\ref{fig:scan}, whereas 
the differential shape of $f^{ }_{k_T}$ at
$T = T^{ }_* = 1$~MeV can be inferred from 
\figs\ref{fig:distribution} and \ref{fig:distribution2}.
The initial neutrino densities yielding the correct dark matter
abundances in all cases (a)-(j)
are summarized in table~\ref{tab_crit}.
It is remarkable that despite quite different asymmetries
(cf.\ table~\ref{tab_crit}), cases (a), (b), (e), (f), (h) and (i) 
produce very similar spectra 
(cf.\ \figs\ref{fig:distribution}, \ref{fig:distribution2}).

%
\begin{table}
	\centering
	\begin{tabular}{|c|c|c|c|}
                \hline
                 \multicolumn{4}{|c|}{
      Initial neutrino number density $n^{ }_{\nu_x}/s$ in units of $10^{-6}$
                 } \\[1mm]
		\hline 
                 case
		 &  $\sin^2(2\theta) = 2\times 10^{-11}$ 
                 &  $\sin^2(2\theta) = 7\times 10^{-11}$ 
                 &  $\sin^2(2\theta) = 20\times 10^{-11}$\\[1mm]
		\hline
		a &  14.14  & 12.25 &  10.81 \\
		b &  19.30  & 17.42 &  15.65 \\
		c &  13.47  & 11.60 &  10.69 \\
		d &  19.11  & 17.80 &  17.15 \\
		e &  33.45  & 30.16 &  27.08 \\
		f &  102.77 &  96.49 & 88.72 \\
		g &  100.56 & 96.85  &  94.99 \\
		h &  82.51  & 72.14 &  63.60 \\
		i &   82.34 & 72.13 &   63.75 \\
		j &  30.91  & 28.02  & 26.65 \\
		\hline
	\end{tabular}

	\caption{\small 
  Initial neutrino densities at $T^{}_\rmi{max}=4$~GeV yielding
  the correct dark matter abundance, 
  expressed as $10^6 n^{ }_{\nu_x} / s$, 
  where $x$ denotes the flavour relevant for the cases (a)-(j)
  (cf.\ \se\ref{se:numerics}). For reasons of numerical reproducibility
  more digits have been shown than is the expected 
  theoretical accuracy of our analysis
  (errors are expected on the $10-20$\% level, 
  mainly from hadronic uncertainties). 
  }
	\label{tab_crit}
\end{table}
%

\begin{figure}[t]
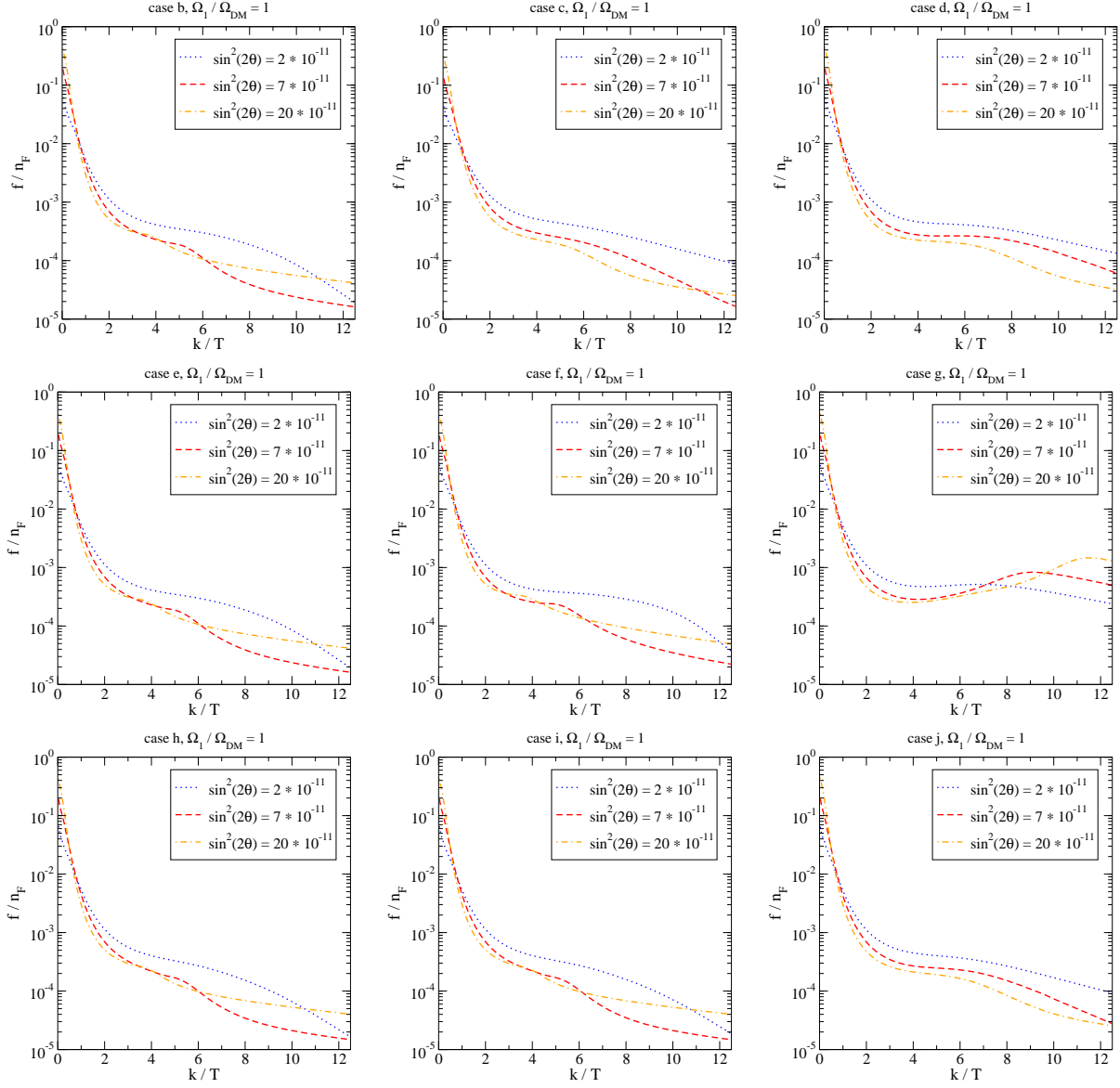



\centerline{%
 \epsfysize=5.0cm\epsfbox{distributionb.eps}%
~~~~\epsfysize=5.0cm\epsfbox{distributionc.eps}%
~~~~\epsfysize=5.0cm\epsfbox{distributiond.eps}%
}

\vspace*{2mm}

\centerline{%
 \epsfysize=5.0cm\epsfbox{distributione.eps}%
~~~~\epsfysize=5.0cm\epsfbox{distributionf.eps}%
~~~~\epsfysize=5.0cm\epsfbox{distributiong.eps}
}

\vspace*{2mm}

\centerline{%
 \epsfysize=5.0cm\epsfbox{distributionh.eps}%
~~~~\epsfysize=5.0cm\epsfbox{distributioni.eps}%
~~~~\epsfysize=5.0cm\epsfbox{distributionj.eps}
}

\caption[a]{\small
 Like \fig\ref{fig:distribution} but for the other cases
 [for ease of comparison, case (e) is reproduced here]. 
}

\la{fig:distribution2}
\end{figure}

%
\section{Conclusions}
\la{se:concl}

In view of the exciting (if unconsolidated) prospect of 
accounting for dark matter through 7.1~keV sterile neutrinos~\cite{obs1,obs2}, 
the purpose of this paper has been to promote a previously proposed 
quantum field theoretic framework~\cite{shifuller} from a qualitative 
towards a  more quantitative level. 
In order to reach this goal, two types of  
``back reactions'' (i.e.\ non-linearities) 
entering the basic equations have been 
derived from stated assumptions  
(cf.\ \ses\ref{ss:production}, \ref{ss:washout}). The relation
of the lepton densities and lepton chemical potentials entering these
equations has been systematically worked out to leading order in
small chemical potentials (cf.\ \se\ref{ss:relation}). The equations have
been written in a form which separately tracks three different flavours of
non-equilibrated lepton densities (cf.\ \eqs\nr{dotfk_2}, \nr{dotna_2}).
Finally, the equations have been numerically solved ``as is'', without
imposing further model assumptions at this stage. 

In a previous study~\cite{shifuller}, which relied otherwise
on similar approximations as the present one, 
it was assumed that all active flavours are in chemical 
equilibrium, and that both the charged and the neutral leptons carry the
same asymmetry, so that the total initial lepton asymmetry is 
effectively $n^{ }_L = 9 n^{ }_{\nu_e}$. This leads to a large 
coefficient $c$ and correspondingly to a maximally efficient resonant
contribution. In reality, as we have discussed, charged leptons cannot
carry such large asymmetries because of electric charge neutrality. 
Concretely, 
this implies that we need a larger active-sterile mixing angle or 
initial asymmetry for a comparable effect. For instance, for case~(a), 
where we find the initial asymmetry 
$n^{ }_{\nu_e}/s = 12.25\times 10^{-6}$ for 
$\sin^2(2\theta) = 7\times 10^{-11}$, the analysis of
ref.~\cite{shifuller} would have produced the correct dark matter
abundance already with $n^{ }_{\nu_e}/s = 9.05\times 10^{-6}$ for
$\sin^2(2\theta) = 7\times 10^{-11}$, or already for 
$\sin^2(2\theta) = 1.5\times 10^{-11}$ with 
$n^{ }_{\nu_e}/s = 12.25\times 10^{-6}$. In other words, the
difference between ref.~\cite{shifuller} and 
the present work is of order unity. On the logarithmic scale 
of \figs\ref{fig:distribution}, \ref{fig:distribution2} 
the distributions of ref.~\cite{shifuller} do however bear
some similarity with ours, if considered
at the same value of $\sin^2(2\theta)$.

Our numerical results have been presented in \se\ref{se:numerics}.
The final spectra for all the cases considered can also be downloaded from 
{\tt http://www.laine.itp.unibe.ch/dmpheno/}. 
It remains a theoretical challenge to confirm whether some of the 
pre-existing neutrino
asymmetries in table~\ref{tab_crit} can indeed be produced by mechanisms
such as the one described in ref.~\cite{ms2}. 

Despite the improvements of the present paper, it should be acknowledged
that the solution still contains 
theoretical uncertainties. The reason is that most of the sterile neutrino
production takes place at temperatures of a few hundred MeV 
(cf.\ \fig\ref{fig:fkT}), where hadronic effects play
a significant role. In our work, hadronic effects have been handled
through a phenomenological recipe introduced in ref.~\cite{numsm}, 
which does correctly incorporate the fact that QCD displays a rapid 
but smooth
crossover rather than an actual phase transition. 
Then hadronic uncertainties remain on a level of 
some tens of percent as discussed previously~\cite{shifuller}. 
Eventually, if the sterile neutrino dark matter scenario establishes
itself, many of these uncertainties can be reduced through 
lattice Monte Carlo measurements. As has been outlined 
in ref.~\cite{hadronic} and in the present paper, 
lattice input is needed for the
equation of state, for quark number susceptibilities, and for
mesonic correlation functions in various quantum number channels. 
The first two ingredients would already be available but it is 
not clear whether including lattice input in some places and not
in others would consistently improve on our results.\footnote{%
 It was verified in ref.~\cite{numsm} that by the time Chiral Perturbation
 Theory is applicable, which would permit for an analytic treatment of 
 hadronic effects, hadronic effects are below the 1\% level. 
 }
Nevertheless their gradual inclusion seems to present an 
interesting challenge for future work. 

%
\section*{Acknowledgements}

M.L thanks D.~B\"odeker and M.~Shaposhnikov for helpful discussions. 
This work was partly supported by the Swiss National Science Foundation
(SNF) under grant 200020-155935.

%
\appendix
\renewcommand{\thesection}{Appendix~\Alph{section}}
\renewcommand{\thesubsection}{\Alph{section}.\arabic{subsection}}
\renewcommand{\theequation}{\Alph{section}.\arabic{equation}}

%
\section{Practical treatment of resonance contributions}

In this appendix we sketch one example for 
how the resonance contributions, outlined
in \se\ref{ss:resonance}, can be handled in practice. 

In general, the resonances cannot be considered as infinitely narrow, 
and their treatment cannot be separated from that of non-resonant
contributions. 
Consider the vicinity of a structure like in \eq\nr{Ra_res}.
Numerical integrations take place on a discrete grid. Suppose that between 
two grid points, $x^{ }_{i-1}$ and $x^{ }_{i}$,
where $x$ can be chosen as $\ln(T^{ }_*/T)$ in \eq\nr{dotfk_2} and
as $E_1$ on the right-hand side of \eq\nr{dotna_2} 
($E^{ }_1 = \sqrt{k_T^2 + M_1^2}$), 
we find that $\mathcal{F}$ has changed its sign; let $x^{ }_0$
denote the location of the zero. 
In the case of the integral on the right-hand side of \eq\nr{dotna_2},
the correct contribution would be
\ba
 \Delta & \approx & 
 \int_{x^{ }_{i-1}}^{x^{ }_{i}} \! {\rm d}x \, 
 \phi(x^{ }_0) 
 \frac{I^{ }_Q(x^{ }_0)}
 { [\mathcal{F}'(x^{ }_0)(x-x^{ }_0)]^2 + I_Q^2(x^{ }_0)}
 \nn 
 & = & 
 \frac{\phi(x^{ }_0)}{|\mathcal{F}'(x^{ }_0)|}
 \biggl\{ \arctan\biggl[ 
 \frac{|\mathcal{F}'(x^{ }_0)|(x^{ }_0
 - x^{ }_{i-1})}{I^{ }_Q(x^{ }_0)} \biggr]
 + 
 \arctan\biggl[
 \frac{|\mathcal{F}'(x^{ }_0)|(x^{ }_{i} - x^{ }_0 )}{I^{ }_Q(x^{ }_0)} \biggr]
 \biggr\}
 \;, \la{Delta}
\ea
whereas naively we could have estimated this through 
\ba
 \delta & \approx & 
 \frac{x^{}_{i} - x^{ }_{i-1}}{2}
 \biggl\{ 
   \frac{\phi(x^{ }_{i-1})I^{ }_Q(x^{ }_{i-1})}
        {\mathcal{F}^2(x^{ }_{i-1}) + I^2_Q(x^{ }_{i-1}) }
  +\frac{\phi(x^{ }_{i})I^{ }_Q(x^{ }_{i})}
        {\mathcal{F}^2(x^{ }_{i}) + I^2_Q(x^{ }_{i}) } 
 \biggr\} 
 \;.
\ea
In order to correct for the error, we may subtract $\delta$ and add $\Delta$
to the result. If $x^{ }_0$ is close to $x^{ }_i$ or $x^{ }_{i-1}$, 
neighbouring cells need to be corrected as well. 

In the case of \eq\nr{dotfk_2}, the integrated result has the structure
\ba
 && \hspace*{-5mm}
 f(x^{ }_i) \; = \;  f(x^{ }_{i-1}) + 
 \sum_a 
 \int_{x^{ }_{i-1}}^{x^{ }_i} \! {\rm d}x' \, 
 \Bigl[ \tilde{\phi}^{ }_a(x') - f(x') \tilde{\chi}^{ }_a(x') \Bigr]
  \la{nonlin} \\
 & \approx & 
 f(x^{ }_{i-1}) + \sum_{a'} \frac{x^{ }_i - x^{ }_{i-1}}{2}
 \Bigl[ 
  \tilde{\phi}^{ }_{a'}(x^{ }_{i-1})
 - f(x^{ }_{i-1}) \tilde{\chi}^{ }_{a'}(x^{ }_{i-1})
 +  \tilde{\phi}^{ }_{a'}(x^{ }_i) 
 - f(x^{ }_{i}) \tilde{\chi}^{ }_{a'}(x^{ }_i)
  \Bigr] \nn 
 & + &    
 \frac{\phi^{ }_r(x^{ }_0)
 - f(x^{ }_0)\chi^{ }_r(x^{ }_0)}{|\mathcal{F}'(x^{ }_0)|}
 \biggl\{ \arctan\biggl[ 
 \frac{|\mathcal{F}'(x^{ }_0)|(x^{ }_0
 - x^{ }_{i-1})}{I^{ }_Q(x^{ }_0)} \biggr]
 + 
 \arctan\biggl[
 \frac{|\mathcal{F}'(x^{ }_0)|(x^{ }_{i} - x^{ }_0 )}{I^{ }_Q(x^{ }_0)} \biggr]
 \biggr\}
 \;, 
 \nonumber
\ea
where 
$a'$ enumerates non-resonant terms;
$r$ is the resonant contribution; 
$\tilde\phi \equiv \phi\, I^{ }_Q / (\mathcal{F}^2 + I_Q^2)$; 
and 
$\tilde\chi \equiv \chi\, I^{ }_Q / (\mathcal{F}^2 + I_Q^2)$. 
Assuming the non-resonant contribution to be
subleading, which can be arranged by choosing 
$x^{ }_i - x^{ }_{i-1}$ sufficiently small, 
the unknown value $f(x^{ }_0)$ can be estimated from 
\be
 \frac{f(x^{ }_{i}) - f(x^{ }_0)}{f(x^{ }_0) - f(x^{ }_{i-1})}
 = 
 \frac{
 \arctan\biggl[
 \frac{|\mathcal{F}'(x^{ }_0)|(x^{ }_{i}
 - x^{ }_0 )}{I^{ }_Q(x^{ }_0)} \biggr]
 }{
 \arctan\biggl[ 
 \frac{|\mathcal{F}'(x^{ }_0)|(x^{ }_0
  - x^{ }_{i-1})}{I^{ }_Q(x^{ }_0)} \biggr]
 }
 \;.
\ee
This implies that we can write
\ba
 && \hspace*{-1cm}
 f(x^{ }_0) \biggl\{ 
 \arctan\biggl[ 
 \frac{|\mathcal{F}'(x^{ }_0)|(x^{ }_0
  - x^{ }_{i-1})}{I^{ }_Q(x^{ }_0)} \biggr]
   + 
 \arctan\biggl[
 \frac{|\mathcal{F}'(x^{ }_0)|(x^{ }_{i}
  - x^{ }_0 )}{I^{ }_Q(x^{ }_0)} \biggr]
 \biggr\} 
 \nn 
 & = & 
 f(x^{ }_{i}) \, 
 \arctan\biggl[ 
 \frac{|\mathcal{F}'(x^{ }_0)|(x^{ }_0
  - x^{ }_{i-1})}{I^{ }_Q(x^{ }_0)} \biggr]
   + 
 f(x^{ }_{i-1})\, \arctan\biggl[
 \frac{|\mathcal{F}'(x^{ }_0)|(x^{ }_{i}
 - x^{ }_0 )}{I^{ }_Q(x^{ }_0)} \biggr]
 \;.
\ea
Inserting this into \eq\nr{nonlin} we can solve for $f(x^{ }_{i})$ 
in terms of the known $f(x^{ }_{i-1})$, including now the non-resonant
contributions as well. 

%

\end{document}